\documentclass[]{amsart}

\usepackage{amsmath}
\usepackage{amssymb}
\usepackage{subcaption}
\usepackage[justification=centering]{caption}
\usepackage{amsthm}
\usepackage{mathtools}
\usepackage{enumitem}

\usepackage{natbib} 

\usepackage{geometry} 
\usepackage[colorlinks]{hyperref} 
\hypersetup{
	linkcolor=blue,
	citecolor=blue,
}
\usepackage[english]{cleveref}

\DeclareMathOperator*{\argmin}{arg\,min}

\newcommand{\diffd}{\mathrm{d}}
\def\diag{\mathop{\rm diag}\nolimits}%
\DeclareMathOperator{\KL}{KL}
\DeclareMathOperator{\tr}{tr}
\DeclareMathOperator{\op}{op}
\DeclareMathOperator{\Id}{Id}
\DeclareMathOperator{\proj}{proj}
\DeclareMathOperator{\refe}{ref}

\newcommand{\ind}{\mathrel{\perp\mspace{-9mu}\perp}}
\newcommand{\notind}{\mathrel{\,\not\!\ind}}
\newcommand{\dphi}{\diffd\varphi}
\newcommand{\opnorm}[1]{\|#1\|_{\mathrm{op}}}
\newcommand{\margs}{\underline{\Sigma}}
\newcommand{\weights}{\underline{w}}
\newcommand{\Fro}{\operatorname{F}}
\newcommand{\sym}{\operatorname{sym}}

\newcommand{\RR}{\mathbb{R}}
\newcommand{\PP}{\mathbb{P}}
\newcommand{\EE}{\mathbb{E}}

\newcommand{\cD}{\mathcal{D}}
\newcommand{\cZ}{\mathcal{Z}}
\newcommand{\cE}{\mathcal{E}}
\newcommand{\cC}{\mathcal{C}}
\newcommand{\cK}{\mathcal{K}}
\newcommand{\cX}{\mathcal{X}}
\newcommand{\cM}{\mathcal{M}}
\newcommand{\cW}{\mathcal{W}}

\newtheorem{theorem}{Theorem}[section]
\newtheorem{remark}{Remark}[section]
\newtheorem{example}{Example}[section]
\newtheorem{proposition}{Proposition}[section]
\newtheorem{corollary}{Corollary}[section]
\newtheorem{lemma}{Lemma}[section]
\newtheorem{definition}{Definition}[section]
\newtheorem{assumption}{Assumption}[section]

\title{Graphical Modeling from Marginal Information via Optimal Transport}

\author{Ulysse Naepels, Paul Freulon and Victor M. Panaretos}

\begin{document}

\begin{abstract}
A graphical model represents multivariate interactions through conditional independence, but requires jointly observed samples to be fitted to data. 
When individuals cannot be tracked and only marginal observations are available, as with destructive measurements leading to cross-sectional data, the problem of graphical modeling is confounded with that of assembling a joint law. 
We show how simultaneous graphical modeling \emph{and} joint law assembly can be realized via a multi-marginal optimal transport problem with graphical cost. Focusing on Gaussian marginals, we generate a joint law satisfying the Markov properties prescribed by a given graph. 
Within the class of joint distributions respecting the graph and marginals constraints, our construction selects a member by making the interactions set by the graph as strong as possible under a dependence budget measured by relative entropy.
In the empirical setting, where only estimates of the marginals are available, we examine the stability of the problem under perturbations of both the marginals and of the graph weights. The resulting error splits into an estimation term, vanishing with the sample size, and a specification term, which quantifies the cost of a misspecified graph. Finally, we illustrate these findings by numerical experiments.
\end{abstract}
\maketitle
\tableofcontents

\section{Introduction}

\subsection{Graphical modeling from marginals}

Graphical models provide an interpretable framework for modeling multivariate interactions through conditional independence \citep{lauritzen1996graphical}. For Gaussian distributions, conditional independence is encoded by the sparsity pattern of the precision matrix, so graphical model estimation reduces to estimating a sparse inverse covariance matrix \citep{dempster1972covariance}.
When samples are observed jointly, the classical theory of covariance selection applies: constraining the sparsity pattern of the precision matrix allows to build the graphical maximum likelihood estimator. 
However, in many applications only partial or marginal observations are available: this is typically the case for destructive measurements such as single-cell experiments \citep{saelens2019comparison, schiebinger2019optimal, weinreb2018fundamental, wolf2019paga}, or whenever individuals cannot be separately tracked \citep{bar1978tracking,leonard2020initial, ouellette2006quantitative, teo1991neural} leading to cross-sectional data.
In such cases, one observes the marginal distributions at each node and may also know the underlying graph, but the cross-node dependence is never observed. Classical covariance selection is therefore inapplicable and the joint distribution can only be partially identified, yielding a class of admissible distributions. This raises a modeling question: given the marginals and a graph, how should one select a joint distribution? And, an inference question: what statistical guarantees hold for the joint distribution constructed from estimated marginals?

We address these questions using entropic optimal transport \citep{cuturi2013sinkhorn}. Given marginals and a graph, we select the joint distribution as the unique optimal coupling of the graph notes subject to an entropic penalty on the induced joint law. The entropic penalty measures the relative distance from the given coupling to the independent one, and acts as a \emph{dependence budget}. The selection spends this budget to make relations along the graph edges as strong as possible but otherwise creates no additional direct interactions. Consequently, the selected coupling matches the marginals and is Markov with respect to the prescribed graph. For example, when marginals represent cross-sections of random trajectories, edges join consecutive times and interactions are attractive: the dependence budget limits the roughness of the reconstructed paths.

The link between entropic transport with graphical cost and graphical models was first pointed out and studied computationally for measures on finite spaces in \citet{Haasler2021multi}. We extend this correspondence to the setting of Gaussian graphical models for which we are able to provide a fairly complete theory. First, for Gaussian marginals, the selected coupling is unique and has an explicit precision matrix. The dependence budget gives the selection a justification (Section \ref{sec:multimarge_ot}). We also investigate what the construction identifies. Specifically, the marginals and the graph do not identify a joint law but only a class of them, and we show this class to be the image of the edge weights under the solution map of the transport problem: the transport formulation amounts to a selection (Section \ref{sec:identifiability}). Moreover, we propose a perturbation theory for the selected coupling in its inputs, the marginals and the weights. We show in particular that they contribute in orthogonal directions (Section \ref{sec:stability}). We further derive a non-asymptotic error bound, a central limit theorem and optimal convergence rates for a plug-in estimator with explicit constants revealing the dependence in the graph, dimensions, and regularization parameter (Section \ref{sec:statistical-scenario}). Finally, we numerically study the empirical behavior of our method, thus validating the proposed framework.

\subsection{Related work}

\textbf{Gaussian graphical models.}
Graphical models give an interpretable representation of the conditional dependence structure of a multivariate distribution, encoding conditional independence through the absence of edges \citep{dawid1979conditional, whittaker2009graphical, lauritzen1996graphical}. When the distribution is Gaussian, this structure is made explicit at the level of the inverse covariance: two variables are conditionally independent given the others if and only if the corresponding entry of the precision matrix vanishes, so that each non-zero entry corresponds to an edge \citep{dempster1972covariance, speed1986gaussian}. Graph discovery - learning the graphical model of a random vector - is therefore equivalent to recovering the support of the precision matrix, the goal of an extensive high-dimensional literature: neighborhood selection, which regresses each variable on the others by the lasso \citep{meinshausen2006high}, and penalized-likelihood methods such as the graphical lasso \citep{yuan2007model, friedman2008sparse, banerjee2008model} and constrained $\ell_1$ minimization \citep{cai2011constrained}. Conversely, when the graph is known, estimating the covariance under the prescribed zero pattern of the precision is the covariance selection problem \citep{dempster1972covariance, lauritzen1996graphical}, which admits an equivalent characterization as a positive-definite matrix completion: among all completions of the observed entries, the maximum-determinant one is the completion whose inverse vanishes off the graph \citep{grone1984positive}. These approaches require jointly observed samples, from which the on-graph entries of a sample covariance can be formed.  Estimators for incomplete data presuppose that the missingness is random so that the joint distribution remains identifiable \citep{stadler2012missing, kolar2012estimating, loh2011high}. Our setting departs from this assumption: the blocks are observed on disjoint samples and never measured jointly, a structural missingness under which the cross-covariances cannot be identified from the data alone.

\textbf{Multi-marginal optimal transport with graphical cost.} To address the aforementioned structural missingness, we exploit tools from optimal transport theory  \citep{ambrosioGradientflowsmetric2005a,villani2009optimal, santambrogio2015optimal,peyre2019computational}. We consider an arbitrary finite number of marginals, an extension known as multi-marginal optimal transport. Uniqueness of the solution was established in \citet{gangbo1998optimal}; see \citet{pass2015multi} for an overview. With two or more marginals, computing the optimal transport requires solving a minimization problem that can be regularized. Entropic regularization, popularized by \citet{cuturi2013sinkhorn} through the Sinkhorn algorithm, triggered renewed interest in entropic optimal transport, notably in the statistical community \citep{panaretos2020invitation, chewi2025statisticalOT}, where it has been shown to improve statistical properties over the classic problem \citep{mena2019statistical, chizat2020faster, rigollet2025sample}. Entropic optimal transport connects to Schrödinger bridge problems \citep{leonard2014Survey, chen2021StochasticControlLiaisons}, which interpolate between probability measures; multi-marginal problems are discrete-time versions of this formulation with added marginal constraints. This motivates the use of optimal transport to reconstruct trajectories from marginal observations alone \citep{Chen2018StateTracking, lavenant2024toward, chizat2022trajectory, hong2025trajectory}, as with destructive measurements \citep{schiebinger2019optimal} or indistinguishable particles in a swarm.
The input measures are taken to be Gaussian, as in \citet{dandapanthula2025optimal}. This retains non-discrete measures of unbounded support while reducing the problem to vectors and covariance matrices, and permits the use of existing results on Gaussian graphical models \citep{uhler2017gaussian}. The cost function is given a graph structure, as in \citet{pass2021cyclic, pass2023monge}; those papers treat the unregularized problem, whereas the present analysis uses entropic penalties, as in \citet{marino2020optimal}.

The combination of entropic regularization and graph structure in the multi-marginal setting is the starting point of \citet{Haasler2021Control, Haasler2021Schrodinger, Haasler2021multi}, where the correspondence is first established. They consider discrete settings with an algorithmic focus and develop inference algorithms for belief propagation. We extend the correspondence in the continuous setting and tackle statistical questions not considered in the discrete and algorithmic case.
Estimation of the coupling is also studied in \citet{li2025sample} under compact support, which excludes the Gaussian case tackled here.

\bigskip
\textbf{Notation and conventions}

We write $S_d(\mathbb{R})$ for the symmetric matrices acting on $\mathbb{R}^d$, equipped with the Frobenius inner product $\langle \Sigma_1, \Sigma_2 \rangle_{\rm F} := \tr(\Sigma_1 \Sigma_2)$ and the operator norm $\|\cdot\|_{\op}$, and $S_d^{+}(\mathbb{R})$, $S_d^{++}(\mathbb{R})$ for the cones of positive semi-definite and positive definite matrices, also written $\Sigma \succeq 0$ and $\Sigma \succ 0$.  If $A$ is a real matrix, we write $\sym(A) = \frac{1}{2}(A+A^\top)$. 
Throughout, we work on the product space $\mathbb{R}^{d \times p} := (\mathbb{R}^d)^{p}$ and underline its elements, writing $\underline{x} = (x_1, \ldots, x_p)$ and, for tuples of matrices, $\margs = (\Sigma_1, \ldots, \Sigma_p) \in \mathcal{X} := S_d(\mathbb{R})^{p}$, with $\langle \underline{A}, \underline{B} \rangle_{\mathcal{X}} := \sum_{i=1}^{p} \tr(A_i B_i)$.
Symmetric matrices acting on $\mathbb{R}^{d \times p}$ form $S_{dp}(\mathbb{R})$, again with the Frobenius inner product. The two spaces are related by the extraction of diagonal blocks $\diag : S_{dp}(\mathbb{R}) \to \mathcal{X}$ and its adjoint $\diag^{*} : \mathcal{X} \to S_{dp}(\mathbb{R})$, the block-diagonal embedding, so that $\langle \underline{A}, \diag(C) \rangle_{\mathcal{X}} = \langle \diag^{*}(\underline{A}), C \rangle_{\rm F}$ for all $\underline{A} \in \mathcal{X}$, $C \in S_{dp}(\mathbb{R})$.

\section{Selecting a joint law by multi-marginal optimal transport}\label{sec:multimarge_ot}

Let $X_1,\ldots,X_p$ be random vectors in $\RR^d$ with laws $\mu_1, \ldots, \mu_p$ and let $G = (V, E)$ be an undirected graph on $V = \{1, \ldots, p\}$ whose edges encode direct interactions between variables. Assume that only the marginals of each $X_i$ is known, so that the joint law of $(X_1,\ldots,X_p)$ is not identified. We introduce a selection principle to choose one admissible coupling based on a minium-energy principle such that (i) the graph prescribes direct dependencies, (ii) the coupling makes relations along edges as strong as a \emph{dependence budget} allows. This is formalized through a multi-marginal optimal transport problem with graphical cost and entropic penalty taken with respect to the product measure.

\subsection{A minimum-energy principle with marginal constraints}

Our first building block is a graph $G = (\{1, \dots, p\}, E)$ over $\{1,\ldots, p \}$ with edges $E \subset \{\{i,j\}~|~i,j\in \{1,\ldots, p\}\}$. To this graph, we associate the ground cost function $c_G : \mathbb{R}^{d \times p } \rightarrow \mathbb{R}_+$ defined by
\begin{equation}\label{eq:graph_ground_cost}
    c_{G}(x_1,\ldots,x_p) := \sum_{\{i,j\} \in E} \|x_i - x_j\|^2.
\end{equation}
This cost is in effect the energy of the joint configuration $(x_1, \dots, x_p)$. 
As we are searching for a coupling, we lift this function \eqref{eq:graph_ground_cost} to $\mathcal{P}(\mathbb{R}^{d \times p})$ through the following functional, defined at $\pi \in \mathcal{P}(\mathbb{R}^{d \times p})$ by
\begin{equation}\label{eq:unregularized_objective}
    \pi \mapsto  \int_{\mathbb{R}^{d\times p}} c_G(x_1,\ldots, x_p)\pi(\diffd x_1,\ldots, \diffd x_p) =\sum_{\{i,j\} \in E } \int_{\mathbb{R}^{d\times p}} \|x_i-x_j\|^2 \pi(\diffd x_1 \cdots \diffd x_p).
\end{equation}
A minimum-energy principle only would lead us to search for minima of \eqref{eq:unregularized_objective}. However, we want our coupling to match the observed marginals $\mu_1,\ldots, \mu_p$. This restriction leads us to optimize over the constraint set
\begin{equation}\label{eq:constraint_multimarge_measure}
	\Pi(\underline{\mu}) := \{\pi \in \mathcal{P}(\mathbb{R}^{d\times p})~|~ \forall j \in \{1,\ldots, p\},~ \proj_j \# \pi = \mu_j \}.
\end{equation}
Combining \eqref{eq:graph_ground_cost}, \eqref{eq:unregularized_objective}, and \eqref{eq:constraint_multimarge_measure} gives the multi-marginal problem
\begin{equation}\label{eq:graph_optimal_transport}
	W_{G}(\underline{\mu}) =  \min_{\pi \in \Pi(\underline{\mu})}  \sum_{\{i,j\} \in E } \int_{\mathbb{R}^{d\times p}} \|x_i-x_j\|^2 \pi(\diffd x_1 \cdots \diffd x_p).
\end{equation}
Problems of the form \eqref{eq:graph_optimal_transport} are well studied in the optimal transport community. We provide reminders for this class of problems in Section \ref{apx:subsec:multimargin-ot} of the Appendix.
Note that each graph gives a specific instance of the multi-marginal optimal transport problem. For instance, choosing  $G$ to be the complete graph on $\{1,\ldots, p\}$ yields the classic multi-marginal problem \eqref{eq:multimarge_ot_euclidean} that we recall in the Appendix. As mentioned in the introduction, this class of problems parameterized by graphs has already been studied by \citet{Haasler2021multi, pass2023monge}.
Our questions are statistical: we focus on which coupling the problem selects among those compatible with the data and how stable it is with respect to the estimated or selected quantities.

We assume the marginal measures to be Gaussian: $\mu_1 = N(0, \Sigma_1), \ldots, \mu_p = N(0, \Sigma_p)$ which yields a reduction to problem \eqref{eq:graph_optimal_transport}. By expanding the squares in the ground cost \eqref{eq:graph_ground_cost}, we get
\begin{equation*}\label{eq:graph_ot_correlation_max}
    W_{G}(\underline{\mu}) = \sum_{i=1}^p \delta_i\mathbb{E}_{\mu_i}[\|X_i\|^2] - 2 \max_{\substack{(X_1,\ldots,X_p) \\ X_i \sim \mu_i}} \sum_{ \{i,j\} \in E} \mathbb{E}[\langle X_i, X_j \rangle],
\end{equation*}
where $\delta_i = \sum_{1 \leq j \leq p} {\bf 1}\{j \sim i\}$ is the degree of vertex $i$. This gives an alternative interpretation to problem \eqref{eq:graph_optimal_transport}. We are searching for a coupling over $\RR^{d\times p}$ that maximizes the correlation between random variables sharing an edge in the graph $G$,  that is, which makes graph-adjacent variables as aligned as possible.
Now, the constraint set consists of the positive semi-definite elements of $S_{dp}(\mathbb{R})$ with diagonal blocks $\Sigma_1,\ldots,\Sigma_p$:
\begin{align}\label{eq:constraint-set}
    \Pi(\margs) := \{ C = (C_{i,j})_{1 \leq i,j \leq p} ~|~ C^T = C,~ C \succeq 0,~\text{and}~ \forall i \in [1,p],~ C_{ii}=\Sigma_i\}.
\end{align}
Given the graph structure of the cost in \eqref{eq:graph_optimal_transport}, we introduce the graph Laplacian $L(G) \in S_p(\mathbb{R})$. This matrix has coefficients $L(G)_{ij} = - {\bf 1}\{i\sim j\}$ if $i \neq j$ and diagonal coefficients $L(G)_{ii}= \delta_i$, the vertex degrees. This standard graph matrix may equivalently be defined via the adjacency matrix, as we recall in Section \ref{app-sec:reminders_graph} of the Appendix. Since the Laplacian $L(G)$ has dimension $p \times p$, to make it match the dimension  $(pd)\times (pd)$ of admissible covariance matrices, we tensorize it with $I_d$, the identity matrix over $\mathbb{R}^d$. That is, $L := L(G)\otimes I_d \in S_{dp}(\mathbb{R})$, a block matrix whose blocks are $L_{ij} = L(G)_{ij}I_d \in S_d(\mathbb{R})$.
We state a reduction of problem \eqref{eq:graph_optimal_transport} to a matrix optimization problem.
\begin{proposition}\label{prop:unregularized-ot-gaussian}
	Let $\mu_1 = N(0, \Sigma_1), \ldots, \mu_p = N(0, \Sigma_p)$ be $p$ centered Gaussian measures defined over $\mathbb{R}^d$. In this case, the graphical optimal transport problem \eqref{eq:graph_optimal_transport} reduces to
\begin{equation}\label{eq:unregularized-ot-gaussian}
    W_G(\underline{\mu}) = \min_{C \in \Pi(\margs)} \langle L, C \rangle_{\rm F}
	\end{equation}
where $L := L(G)\otimes I_d$ is the tensorized Laplacian matrix and $\langle L, C \rangle_{\rm F} = \tr(LC)$. Moreover, if $C^{\star} \in S_{dp}(\mathbb{R})$ is a solution to problem \eqref{eq:unregularized-ot-gaussian}, the Gaussian coupling $N(0,C^{\star})$ is a solution to problem \eqref{eq:graph_optimal_transport}.
\end{proposition}

The constraint set \eqref{eq:constraint-set} is convex and compact, so \eqref{eq:graph_optimal_transport} admits at least one solution. Uniqueness, studied in a more general setting by \citet{pass2021cyclic, pass2023monge}, may fail: by \cite[Prop.~2.1.1]{pass2023monge}, the solution is not unique when $G$ is disconnected. A second obstacle is that the solution is typically deterministic \cite[Thm.~3.1, Thm.~4.1]{pass2023monge}. Both obstacles are removed by entropic regularization, introduced in the next section.

\subsection{The entropic graphical problem}

To avoid non-uniqueness and deterministic solutions, we consider a penalized problem where an entropic penalty is added to the multi-marginal objective function. For $\varepsilon > 0$, we define the coupling as the solution of
\begin{equation}\label{eq:entropy_graph_transport}
	W_{G}^{\varepsilon}(\underline{\mu}) := \min_{\pi \in \Pi(\underline{\mu})}  \sum_{\{i,j\} \in E} \int_{\mathbb{R}^{d\times p}} \|x_i-x_j\|^2 \pi(\diffd x_1 \cdots \diffd x_p) + 2\varepsilon \KL(\pi |\mu_1 \otimes \cdots \otimes \mu_p).
\end{equation}
Here $\KL$ denotes the Kullback-Leibler divergence defined\footnote{We consider the logarithm function to be defined over $[0,+\infty)$, and to have range $\{-\infty\}\cup \mathbb{R}$. At zero, we follow the conventions $\log(0) = -\infty$ and $0\log(0)=0$.} for two measures $\pi$ and $\pi_{\refe}$ by $\KL(\pi|\pi_{\refe}) := \int \log \frac{d\pi}{d\pi_{\refe}}(x) \pi(\diffd x)$ if $\pi$ is absolutely continuous with respect to $\pi_{\refe}$; and $\KL(\pi|\pi_{\refe}):= +\infty$ otherwise.
As in the previous section, we focus on Gaussian measures.

\begin{proposition}\label{prop:entropic-ot-gaussian}
	Let $\mu_1,\ldots, \mu_p$ be $p$ Gaussian measures on $\mathbb{R}^d$ with respective covariance matrices $\Sigma_1,\ldots ,\Sigma_p$. Denoting by $\diag^*(\margs^{-1})$ the block diagonal matrix with blocks $\Sigma_1^{-1}, \ldots, \Sigma_p^{-1}$, the entropic graphical optimal transport problem reduces to
	\begin{equation}\label{eq:entropic-ot-gaussian}
		W_{G}^\varepsilon (\underline{\mu})= \min_{C \in \Pi(\margs)} \langle L, C \rangle_{\rm F}   - \varepsilon \log  \det(\diag^*(\margs^{-1}) C),
	\end{equation}
    where $L$ is the tensorized Laplacian matrix.
\end{proposition}
As in Proposition \ref{prop:unregularized-ot-gaussian}, the problem over probability measures reduces to one over matrices, and we identify the solution $C^{\star}$ to \eqref{eq:entropic-ot-gaussian} with the Gaussian coupling $N(0, C^{\star})$ solving \eqref{eq:entropy_graph_transport}. Unlike the unregularized case, this problem is well-posed.

\begin{lemma}\label{lem:uniqueness_entropic}
Let $\Sigma_1,\ldots, \Sigma_p$ be a collection of $p$ full-rank covariance matrices. Problem \eqref{eq:entropic-ot-gaussian} has a unique solution $C^{\star} \in S_{dp}(\mathbb{R})$ which is positive-definite.
\end{lemma}

Lemma \ref{lem:uniqueness_entropic} motivates the addition of the entropic term: it yields a unique, invertible solution, which the unregularized problem does not. Uniqueness also defines the map $\margs \mapsto C_{\margs}$, studied in Section \ref{sec:stability}.
 The entropic term also allows the selection to be explained through a dependence budget.
Indeed, we can show (Appendix \ref{subsec:proof_gauss_condindep}) that the problem admits the following formulation:
\begin{align}\label{eq:dependence-budget}
    W_{G}^\varepsilon (\underline{\mu}) = \min\left\{ \langle L,C\rangle_{\rm F}~|~C\in\Pi(\margs),\ \cD(C) \leq \cD(C^{\star}_{\varepsilon}) \right\} + 2 \varepsilon \cD(C^{\star}_{\varepsilon}),
\end{align}
where $\cD(C)=\KL (N(0,C)\,|\,\mu_1\otimes \dots \otimes\mu_p )$ is the total correlation \citep{watanabe1960information} of a covariance matrix and encodes the total dependence of a coupling. In particular, it is null under independence and diverges for degenerating couplings. The reformulation states that the selected coupling makes graph-adjacent variables as dependent as possible \emph{under a given total dependence budget}.

\subsection{Characterization of the solution and conditional independence}

Let $(X_1, \ldots, X_p)$ be a Gaussian vector whose covariance solves \eqref{eq:entropic-ot-gaussian}. If $G$ is disconnected, then variables indexed by vertices in distinct connected components are independent. Beyond this, the solution satisfies a \textit{conditional} independence property, which we obtain by characterizing the inverse of the solution to \eqref{eq:entropic-ot-gaussian}.

\begin{theorem}\label{thm:graph_transport_optimality_condition}
    Let $\Sigma_1,\ldots,\Sigma_p$ be $p$ covariance matrices defining $p$ centered Gaussian measures. Then, there exists a vector $\underline{\Lambda}$ such that the entropic optimal transport has solution given by the Gaussian measure with covariance $C^{\star} \in S_{pd}(\mathbb{R})$ satisfying
    \begin{equation}\label{eq:kkt_conditions}
        \left\{\begin{array}{rcl}
            C^{\star} &=& \varepsilon[L+\diag^*(\underline{\Lambda})]^{-1} \\
            \diag(C^{\star}) & = & \margs\\
        \end{array}\right.
    \end{equation}
    where $L$ is the tensorized Laplacian matrix.
\end{theorem}
Characterization \eqref{eq:kkt_conditions} follows from the Karush--Kuhn--Tucker conditions, in which $\underline{\Lambda} \in S_d(\mathbb{R})^p$ solves the dual problem; since the dual serves only to describe the primal solution, we defer it to Section \ref{sec:dual_problem}.
Theorem \ref{thm:graph_transport_optimality_condition} shows that the off-graph entries of $(C^\star)^{-1}$ vanish. Combined with the characterization of conditional independence for Gaussian vectors, this gives the following.

\begin{corollary}\label{cor:conditional_indep_eot}
If $\mu_1,\ldots, \mu_p$ are Gaussian measures all with full rank covariances, then the solution to the problem \eqref{eq:entropy_graph_transport} is a Gaussian measure $N(0,C^{\star})$ whose conditional independence graph is $G$.
\end{corollary}
Corollary \ref{cor:conditional_indep_eot} shows that the graph acts on the joint distribution through conditional independence (and, equivalently for Gaussian distributions, through the inverse covariance). The discrete counterpart of this result is due to \citet{Haasler2021multi}. It therefore offers an alternative to construct a covariance with prescribed conditional independence structure: specify the graph in the multi-marginal problem.

\begin{example}[Banded graph]\label{ex:banded_graph}
    We consider the family of banded graphs as a recurring example throughout the paper. For an integer $b \geq 1$, we call $b$-banded graph, the graph $G$ with edges $E = \{ \{i,j\}~:~ 0<|i-j| \leq b\}$. Its adjacency matrix has coefficients $A_{ij}(G) = {\bf 1}\{0<|i-j| \leq b\}$. While relatively simple, the family of banded graphs describes various levels of connectivity depending on the chosen band $b$. This family of graphs enables us to interpolate between the path graph (by choosing $b=1$) and the complete graph (by choosing $b=p-1$). Section \ref{sec:stability} will show that the maximum degree of the graph plays a key role in stability results. For a $b$-banded graph with small $b$, its maximum degree is given by $\delta_{\max} = 2b$.
\begin{figure}[h!]
    \centering
    \includegraphics[width=0.7\textwidth]{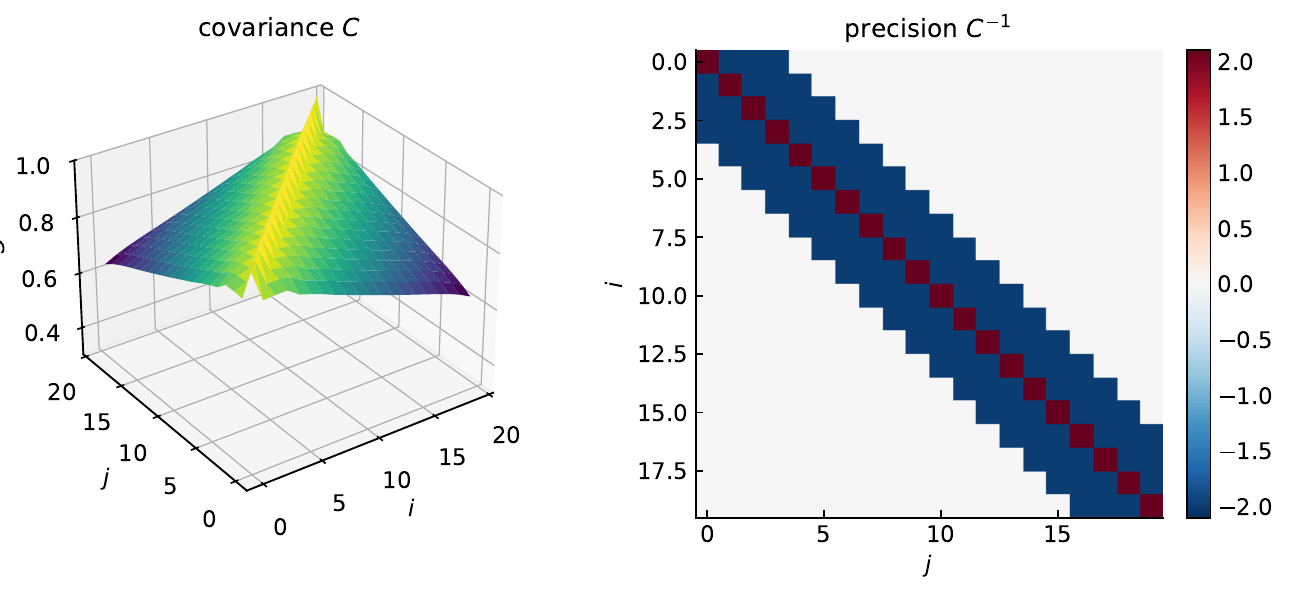}
    \centering
    \caption{Covariance (left) and precision (right) matrices for a $3$-banded graph with $p=20$, $\varepsilon=0.5$ and unit marginals.}
    \label{fig:banded-graph-example}
\end{figure}
\end{example}

Applying Corollary \ref{cor:conditional_indep_eot} to the $b$-banded graph gives the following. Let $C^{\star}$ solve \eqref{eq:entropic-ot-gaussian} with $G$ the $b$-banded graph. If $(X_1, \ldots, X_p) \sim N(0, C^{\star})$, then for every pair $(i,j)$,
\begin{equation*}
        X_i \ind X_j ~|~ \{X_1,\ldots,X_p\}\setminus\{X_i,X_j\} \quad \Longleftrightarrow \quad |i-j| > b.
\end{equation*}

\subsection{Weighted and signed interactions}\label{subsec:weighted}

The graphical cost \eqref{eq:graph_ground_cost} gives the same importance to all edges and rewards alignment. When $X_j\equiv X(t_j)$, $j=1,...,p$, represents the trajectory of a particle in $\mathbb{R}^d$, this is natural: consecutive positions should be as correlated as possible so that paths are not erratic. We can however consider situations where edge weights vary, and where connected states are repulsive rather than attractive: two adjacent variables may be expected to move in opposite directions. We therefore consider a general graph where each pair $i \neq j$ carry a matrix weight $w_{ij} \in M_d(\RR)$ such that $w_{ji} = w_{ij}^\top$ and denote by $\cW_d$ the linear space of such weights $\weights = (w_{ij})_{i\neq j}$.

Because the marginals are fixed, only cross terms influence the quadratic cost. Writing $w_{ij} =A_{ij}^\top B_{ij}$ (e.g. with a singular value decomposition), we consider the cost
\begin{align*}
    c_{\weights}(x_1,\ldots,x_p)& =\sum_{i<j}\|A_{ij}x_i-B_{ij}x_j\|^{2} \\
    &=\sum_{i<j}\bigl(\|A_{ij}x_i\|^{2}+\|B_{ij}x_j\|^{2}\bigr)-2\sum_{i<j}x_i^{\top}w_{ij}\,x_j.
\end{align*}
The first sum integrates to a constant over $\Pi(\margs)$, so that the solutions do not depend on the factorization $w_{ij}=A_{ij}^\top B_{ij}$ but only on the cross term in $\weights$. 
For example, scalar weights correspond to an attraction or repulsion based on their signs, and orthogonal weights reward alignment up to a rotation. Hence, the problem also reduces to 
\begin{align}\label{eq:entropic-ot-weighted}
    \min_{C\in\Pi(\margs)}\ \langle L_{\weights},C\rangle_{\Fro}-\varepsilon\log\det C,
\end{align}
where $L_{\weights}\in S_{dp}(\RR)$ has off-diagonal blocks 
$(L_{\weights})_{ij}=-w_{ij}$ and diagonal blocks 
$(L_{\weights})_{ii}=\sum_{j\neq i} \sym(w_{ij})$.
Moreover, the characterization in \Cref{thm:graph_transport_optimality_condition} still holds, i.e.
\begin{align*}
    \varepsilon C_{\weights}^{-1}=L_{\weights}+\diag^*(\underline{\Lambda}),
    \qquad \diag(C_{\weights})=\margs ,
\end{align*}
and the conditional independence graph of $C_{\weights}$ is given by $\{\{i,j\}: w_{ij} \neq 0\}$. The weights thus encode a prior choice of the sign and strength of each direct interaction. Unless stated otherwise, we use the default weights $\mathbf{1}_E\otimes I_d$; the results of \Cref{sec:stability} and \Cref{sec:statistical-scenario} hold for all $\weights\in\cW_d$.

\begin{remark}
     Taking repulsive weights $w_{k,k+1} = -I_d$ for the path graph creates instability: it amounts to selecting the law of $((-1)^{k} Y_k)$ where $(Y_k)_k$ is the default transport coupling. Consecutive value are then negatively correlated by repulsions, and the constructed path oscillates around the mean. That's also why choosing attractive weights is more natural for trajectories.
\end{remark}

\section{Partial identification and selection}\label{sec:identifiability}

The constraints imposed on the graph and the marginal distributions allow us to identify a unique \emph{class} of joint distributions, from which the entropy transport problem selects one. We show that the weighted transport problem fully parametrizes this class, and bound its identifiability radius.

\subsection{Identified set and selection rule}

Fix $\varepsilon>0$ and $\margs$ with $\Sigma_i\in S_d^{++}(\RR)$, and denote by
$C^{\star}$ the solution of \eqref{eq:entropic-ot-gaussian}. For a graph $G=(V,E)$ and marginals $\margs$, we define the \emph{graphical class associated with $G$ and $\margs$} as
\begin{align}\label{eq:model-class}
    \cM(G, \margs) = \Bigl\{C\in S_{dp}^{++}(\RR):
    \diag(C)=\margs\ \text{ and }\ (C^{-1})_{ij}=0
    \ \text{ for all } i\neq j \text{ with } \{i,j\}\notin E\Bigr\}.
\end{align}
$\cM(G, \margs)$ is the set of covariances of non-degenerate centered Gaussian vectors that are Markov with respect to $G$.
By \Cref{thm:graph_transport_optimality_condition} we have
$(C^{\star})^{-1}_{ij}=-\varepsilon^{-1}{\bf 1}\{i\sim j\}I_d$, so $C^{\star}\in\cM(G, \margs)$ and the independent coupling $\diag^*(\margs)\in\cM(G, \margs)$.
Model \eqref{eq:model-class} corresponds to the set of transport solutions with varying weights.
We can hence define a solution map which we characterize below.
\begin{proposition}\label{prop:identifiability}
    Let $\Sigma_1,\ldots,\Sigma_p\in S_d^{++}(\RR)$ and $\varepsilon>0$.
    The solution map $\chi:\weights\mapsto C_{\weights}$ is a
    bijection from $\cW_d$ onto $\Pi(\margs)\cap S_{dp}^{++}(\RR)$,
    with inverse $\iota(C)_{ij}=-\varepsilon(C^{-1})_{ij}$, and it restricts to a
    bijection
    \begin{equation}\label{eq:chi-bijection}
        \chi:\ \bigl\{\weights\in\cW_d:\ w_{ij}=0\ \text{for }\{i,j\}\notin E\bigr\}
        \ \xrightarrow{\ \sim\ }\ \cM(G, \margs).
    \end{equation}
    Moreover $\chi$ depends on $(\weights,\varepsilon)$ only through
    $\weights/\varepsilon$.
\end{proposition}

This result has two consequences.
\begin{itemize}[nosep]
    \item The model class corresponds to the standard Gaussian graphical models.
    The unweighted problem \eqref{eq:entropic-ot-gaussian} selects $\weights={\bf 1}_E\otimes I_d$, i.e. all edges carry the same partial precision. The identifiable set has dimension $\dim\cM_G=|E| d^2$.
    \item The sign and strength of the interactions are not identified: for any $\weights \in \cW_d$, the selected joint distribution is compatible with the observations. Moreover, Since $\chi$ depends only on $\weights/\varepsilon$, the regularization level and the overall scale of the weights are the same parameter.
\end{itemize}

\subsection{The identifiability radius}

We established that the marginals and the graph do not determine a joint law. Therefore, any selection carries a modeling error. In order to bound this error, we measure the diameter of the identified set of admissible couplings.
Specifically, we bound the worst case bias, or identifiability radius
\begin{align*}
    B_{\varepsilon}(G,\margs)
     = \sup_{C\in\cM(G,\margs)}\|C-C^{\star}\|_{F}.
\end{align*}
The following result shows that the radius is controlled by the sizes of the connected components of $G$ alone, and is at most of order $\sigma^2_{\max}p\sqrt{d}$ for every connected graph, uniformly in the level of regularization.
\begin{proposition}\label{prop:bias-upper}
    Let $V_1,\ldots,V_m$ be the connected components of $G$, of cardinalities
    $p_1,\ldots,p_m$. Then $B_{\varepsilon}(G,\margs)$ is
    finite and
    \begin{equation}\label{eq:bias-upper}
        \Bigl(\sum_{i\neq j}\|C^{\star}_{ij}\|_{\Fro}^{2}\Bigr)^{1/2} \leq 
        B_{\varepsilon}(G,\margs)
        \ \leq\ 2\sigma^2_{\max}\sqrt{d\textstyle\sum_{c}p_c(p_c-1)}.
    \end{equation}
    In particular, if $G$ is connected,
    $B_{\varepsilon}(G,\margs)\leq2\sigma^2_{\max}\sqrt{d\,p(p-1)}$.
\end{proposition}

\section{Sensitivity of the solution to the marginals and the graph weights}\label{sec:stability}

The selection rule has two inputs, the marginals $\margs$ and the graph weights $\weights$. This section studies and quantifies how the selected coupling responds to input perturbations. Specifically, we prove that the mappings sending the marginals (resp. the graph weights) to the solution are well-defined, injective, and differentiable.
Such stability questions are also typical of optimal transport \citep{letrouit2025quantitative}.

\subsection{Sensitivity to the marginals}\label{subsec:stability_marginals}

In this subsection, the weights are fixed: for brevity we consider the default weights, but the results can be extended to every $\weights \in \cW_d$. 
We first show that the solution map is smooth in the marginals, and then bound its differential, which is the quantity governing the propagation of the marginals estimation error.
For $C\in S_{dp}^{++}(\RR)$, let $\Phi_C:S_d(\RR)^{p}\to S_d(\RR)^{p}$ be
\begin{equation}\label{eq:Phi-def}
\Phi_C(\underline V):=\diag \bigl(C\,\diag^{*}(\underline V)\,C\bigr),
\qquad\text{i.e.}\qquad
[\Phi_C(\underline V)]_i=\sum_{k=1}^{p} C_{ik}\,V_k\,C_{ik}^{\top}.
\end{equation}
It is self-adjoint and positive definite:
for every $\underline{V}\in S_d(\RR)^{p}$, $\langle \underline{V}, \Phi_C(\underline{V})\rangle_{\cX} = \| C^{1/2} \diag^*(\underline{V})C^{1/2} \|^2_{\Fro},$
which vanishes only if $\underline{V}=0$. In particular $\Phi_C$ is invertible.

\begin{theorem}\label{thm:regularity-implicit}
	The map $\varphi: \margs \in (S_d^{++}(\RR))^{p} \mapsto C_{\margs} \in S_{dp}^{++}(\RR)$, where $C_{\margs}$ is the solution of \eqref{eq:entropic-ot-gaussian}, is well defined, injective, and continuously differentiable on all of
	$(S_d^{++}(\RR))^{p}$, with differential
	\begin{equation}\label{eq:diff-exact}
	\dphi(\underline{H}) = C \diag^*\bigl(\Phi_C^{-1} (\underline{H})\bigr)C,
	\qquad \underline{H} \in S_d(\RR)^p,\quad C:=C_{\margs} .
	\end{equation}
\end{theorem}

\begin{proof}
Fix $\margs$ a vector of $p$ marginal covariances $\Sigma_i \in S_d^{++}(\RR)$, and write $C = C_{\margs}$ for the solution of \eqref{eq:entropic-ot-gaussian}.
Introducing the optimal dual variable $\underline{\Lambda}$, the Karush--Kuhn--Tucker conditions are
\begin{equation}\label{eq:kkt_conditions_bis}
	L -\varepsilon C^{-1}  + \diag^*(\underline{\Lambda})= 0, \qquad \diag(C)- \margs = 0.
\end{equation}
Let $\cX^{++}:=(S_d^{++}(\RR))^{p}$ and $\cZ^{++}:=S_{dp}^{++}(\RR)\times S_d(\RR)^{p}$, and rewrite \eqref{eq:kkt_conditions_bis} as $f(\margs,(C,\underline{\Lambda})) = 0$ where $f$ is defined on the open set $\cX^{++}\times\cZ^{++}$ by
\begin{equation}\label{eq:equation_implicit}
f(\margs,(C,\underline{\Lambda})) := (L -\varepsilon C^{-1}  + \diag^*(\underline{\Lambda}), \diag(C)- \margs).
\end{equation}
By \Cref{lem:uniqueness_entropic}, $C$ is invertible, so $f$ is differentiable at $(\margs, C, \underline{\Lambda})$, with
\begin{equation}\label{eq:differential_equation_implicit}
    \diffd f(\underline{H}, K, \underline{V}) = (\diag^*(\underline{V}) +\varepsilon C^{-1} K C^{-1}, \diag(K) - \underline{H}).
\end{equation}
We check that $\diffd f|_{\cZ}$ is invertible. We have $(K, \underline{V}) \in \ker\{\diffd f|_{\cZ}\}$ if and only if $\varepsilon C^{-1}K C^{-1} + \diag^*(\underline{V})= 0$ and $\diag(K)=0$, which gives $\Phi_C(\underline{V})=\diag(C\diag^*(\underline{V})C)=-\varepsilon\diag(K)=0$ and hence $\underline{V}=0$ because $\Phi_C$ is invertible. It follows that $\diffd f|_{\cZ}$ is bijective.

By the implicit function theorem, $\varphi$ is $\cC^1$ in a neighborhood of $\margs$. Since $\margs$ is arbitrary in the open set $\cX^{++}$, we get that $\varphi$ is $\cC^1$ globally. Injectivity is immediate from $\diag\circ\varphi=\Id$.
The implicit function theorem gives $\diffd\varphi=-(\diffd f|_{\cZ})^{-1}\circ\diffd f|_{\cX}$ on the first component, and $\diffd f|_{\cX}(\underline{H})=(0,-\underline{H})$, so $(K,\underline{V})=\diffd(\varphi,\psi)_{\margs}(\underline{H})$ solves $\varepsilon C^{-1}KC^{-1}+\diag^*(\underline{V})=0$ and $\diag(K)=\underline{H}$. The first equation gives $K=-\varepsilon^{-1}C\diag^*(\underline{V})C$; taking diagonal blocks and using the second we get $\Phi_C(\underline{V})=-\varepsilon\underline{H}$, that is $\underline{V}=-\varepsilon\Phi_C^{-1}(\underline{H})$.
\eqref{eq:diff-exact} follows.

\end{proof}

The differential is driven by two spectral quantities: the size of $C$ and variations compatible with the marginal constraints, measured by $\Phi_C$. We write
$\delta_{\max}:=\max_i\delta_i$ for the maximum degree of $G$ (in the case of varying weights, we take $\delta_{\max}:=\max_{i} \sum_{j \neq i} \opnorm{ w_{ij}}$).

\begin{proposition}\label{prop:bound-diff}
    Let $C=C_{\margs}$ solve \eqref{eq:entropic-ot-gaussian} and let
    $\dphi=\diffd\varphi_{\margs}$. Suppose
    $\sigma^2_{\min}I_d\preceq\Sigma_i\preceq\sigma^2_{\max}I_d$ for every $i$. Then
    \begin{equation}
        \opnorm{\dphi}
        \ \leq\ \frac{\lambda_{\max}(C)}{\bigl(\sigma^2_{\min}\lambda_{\min}(C)\bigr)^{1/2}}.
    \end{equation}
    In particular, in terms of the inputs,
    \begin{equation}\label{eq:bound-diff}
        \opnorm{\dphi}\ \leq\ \frac{\lambda_+}{\sigma^2_{\min}}
        \sqrt{1+\frac{\delta_{\max}(\sigma^2_{\min} + \sigma^2_{\max})}{\varepsilon}},
        \qquad
        \lambda_{+} := \sigma^2_{\max} \min\Bigl\{ p,\ \frac{\varepsilon}{(\varepsilon - \delta_{\max} \sigma^2_{\max})_+} \Bigr\}.
    \end{equation}
\end{proposition}

Note that a strongly correlated coupling has a large $\lambda_{\max}(C)$ and a small $\lambda_{\min}(C)$, which increases the bound twice. Therefore, increasing the correlations (typically with high connectivity and small $\varepsilon$) makes the rule less stable.
The second bound \eqref{eq:bound-diff} splits into two cases.
\begin{itemize}[nosep]
    \item If $\varepsilon-\delta_{\max}\sigma^2_{\max}$ is bounded away from zero, then $\lambda_+$ does not depend on $p$ and $\opnorm{\dphi}=O(1)$ uniformly in $p$. This encompasses bounded-degree graphs if $\varepsilon$ is large enough.
    \item Otherwise $\lambda_+=p\sigma^2_{\max}$ and $\opnorm{\dphi}\lesssim p\sqrt{\delta_{\max}/\varepsilon}$; e.g. for the complete and star graphs, $\delta_{\max}=p-1$ and $\opnorm{\dphi}\lesssim p^{3/2}\varepsilon^{-1/2}$.
\end{itemize}
The upper bound  depends on $\margs$ only through $\sigma^2_{\min}$ and $\sigma^2_{\max}$; \Cref{thm:nonasymptotic} exploits this uniformity to turn \eqref{eq:bound-diff} into a non-asymptotic bound.

\subsection{Sensitivity to the weights}\label{subsec:stability_graph}

We study the stability of the solution with respect to the weights, which encode the graph edge strength and sign. In particular, we quantify how much a misspecified graph impacts the selected coupling.
We equip $\cW_d$ with $\|\underline{h}\|_{\cW_d}^{2}:=\sum_{\{i,j\}\in\cE}\|h_{ij}\|_{\Fro}^{2}$, where $\cE:=\{\{i,j\}:1\leq i<j\leq p\}$.

\begin{theorem}\label{thm:diff-graph}
    The map $\chi:\cW_d\to S_{dp}^{++}(\RR)$ is continuously differentiable, and
    for $\underline{h}\in\cW_d$, $C:=C_{\weights}$ and
    $K:=\diffd\chi_{\weights}(\underline{h})$,
    \begin{equation}\label{eq:diff-exact-graph}
        K=-\frac{1}{\varepsilon}\Bigl[CL_{\underline h}C
        -C\diag^*\bigl(\Phi_C^{-1}\bigl(\diag(CL_{\underline h}C)\bigr)\bigr)C\Bigr],
        \qquad \diag(K)=0 ,
    \end{equation}
    and $\diffd\chi_{\weights}$ is one-to-one. Its operator norm satisfies
    \begin{equation}\label{eq:opnorm-chi}
        \opnorm{\diffd\chi_{\weights}}\ \leq\
        \frac{\sqrt{2}\,\lambda_{\max}(C_{\weights})^{2}}{\varepsilon}
        \ \leq\ \frac{\sqrt{2}\,\lambda_{+}^{2}}{\varepsilon},
    \end{equation}
    with $\lambda_{+}$ as in \Cref{prop:bound-diff}.
\end{theorem}

The bound \eqref{eq:opnorm-chi} shows that the sensitivity comes from $\lambda_{\max}(C_{\weights})^{2}/\varepsilon$, so the two cases of \Cref{prop:bound-diff} also apply here: the constant is uniform in $p$ for bounded weighted degree and large $\varepsilon$, and of order $p^{2}$ otherwise.
The factor $1/\varepsilon$ is exact since by \Cref{prop:identifiability} the solution depends on $(\weights,\varepsilon)$ only through $\weights/\varepsilon$.

\begin{remark}[Graph misspecification]\label{rem:misspec}
    $\cW_d$ is a linear space on which \eqref{eq:opnorm-chi} holds uniformly, so we get that for all $\weights_1,\weights_2\in\cW_d$,
    \begin{equation}\label{eq:lipschitz-graph}
        \bigl\|C_{\weights_1}-C_{\weights_2}\bigr\|_{\Fro}\ \leq\
        \frac{\sqrt{2}\,\lambda_{+}^{2}}{\varepsilon}\,
        \bigl\|\weights_1-\weights_2\bigr\|_{\cW_d},
        \qquad \lambda_{+}\leq p\,\sigma^2_{\max}.
    \end{equation}
    Deleting a set $S\subseteq\cE$ of edges from a correctly specified $\weights^{\star}$ therefore changes the coupling by at most $\sqrt{2}\lambda_{+}^{2}\varepsilon^{-1} (\sum_{e\in S}\|w_e^{\star}\|_{\Fro}^{2})^{1/2}$. Similarly, reversing the sign of interactions on $S$ (that is, changing attractions to repulsions) changes the coupling by at most $2\sqrt{2}\lambda_{+}^{2}\varepsilon^{-1} (\sum_{e\in S}\|w_e^{\star}\|_{\Fro}^{2})^{1/2}$.
\end{remark}

\begin{remark}[Joint differential]\label{rem:joint-differential}
    The two sensitivity analyses can be combined: the joint map $\Theta:(\margs,\weights)\mapsto C_{\margs,\weights}$ is $\cC^{1}$ on $(S_d^{++}(\RR))^{p}\times\cW_d$, with $\diffd\Theta_{(\margs,\weights)}(\underline{H},\underline{h}) =\dphi(\underline{H})+\diffd\chi_{\weights}(\underline{h})$. The two variation sources decomposes additively and do not interact at first order.
\end{remark}

\section{Estimation and specification}\label{sec:statistical-scenario}

The selection rule has two inputs of different nature: the marginals, estimated from data, and the graph weights, which are specified. 
We therefore treat the two separately. We first bound the estimation error coming from the data, and then bound the specification coming from the selection rule.
\begin{assumption}\label{assum:sampling}
    For each marginal $\mu_i^{\star} = N(0, \Sigma_i^{\star})$, we observe $n_i$ i.i.d. samples $X_i^{(1)},\ldots, X_i^{(n_i)}$ and denote $\widehat{\Sigma}_i$ the sample covariance. Assume $n_i=\lceil\gamma_i n\rceil$ for fixed $\gamma_i>0$, with $\gamma_{\min}=\min_i\gamma_i$.
\end{assumption}
We carry out the analysis with the sample covariance, but the method is plug-in: any marginal estimator could be used.
With $G$ given, we estimate $C^\star$ by plug-in:
\begin{equation}\label{eq:empirical_got}
    \widehat{C}_n := \argmin_{C \in \Pi(\widehat{\Sigma}_1,\ldots, \widehat{\Sigma}_p)}
    \langle L , C \rangle_{\Fro}
    - \varepsilon \log \det\bigl(\diag^*(\widehat{\margs}^{-1}) C\bigr).
\end{equation}
We write $\sigma^{2}_{\min}:=\min_i\lambda_{\min}(\Sigma_i^{\star})$ and $\sigma^{2}_{\max}:=\max_i\lambda_{\max}(\Sigma_i^{\star})$.

\subsection{Estimation error}\label{subsec:estimation}

The bound of \Cref{prop:bound-diff} constrains $\margs$ only through
$\sigma^2_{\min}$ and $\sigma^2_{\max}$, so it holds uniformly over the convex set
\begin{equation}\label{eq:Kset}
    \cK(\sigma^2_{\min},\sigma^2_{\max}):=\bigl\{\margs\in\cX:\
    \sigma^2_{\min}I_d\preceq\Sigma_i\preceq\sigma^2_{\max}I_d\ \ \forall i\bigr\}.
\end{equation}
Since $\varphi$ is $\cC^{1}$ on all of $\cX^{++}$ by \Cref{thm:regularity-implicit}, we can integrate the infinitesimal bound to obtain a global Lipschitz control.
We then apply it on an event where $\widehat{\Sigma}_i$ is close to $\Sigma_i^{\star}$ to obtain a non-asymptotic error bound.
Throughout we write
\begin{align*}
    K(\varepsilon, p, \delta_{\max}) = L\left(\frac{1}{2}\sigma^2_{\min}, \ 2 \sigma^2_{\max}\right)
\end{align*}
where $L(\sigma^2_{\min},\sigma^2_{\max}) :=\frac{\sigma^2_{\max}}{\sigma^2_{\min}} \min\Bigl\{p,\ \frac{\varepsilon}{(\varepsilon-\delta_{\max}\sigma^2_{\max})_+}\Bigr\} \sqrt{1+\frac{2\delta_{\max}\sigma^2_{\max}}{\varepsilon}}$ is the upper-bound \eqref{eq:bound-diff}.

\begin{theorem}[Non-asymptotic error]\label{thm:nonasymptotic}
    For all $\margs_1,\margs_2\in\cK(\sigma^2_{\min},\sigma^2_{\max})$,
    \begin{equation}\label{eq:global-lipschitz}
        \|C_{\margs_1}-C_{\margs_2}\|_{\Fro}
        \ \leq\ L(\sigma^2_{\min}, \sigma^2_{\max})\,
        \|\margs_1-\margs_2\|_{\cX}.
    \end{equation}
    Under assumption \ref{assum:sampling}, let $\widehat{C}_n$ be the estimator \eqref{eq:empirical_got}. Then, there is a constant  $c_0 > 0$ such that for every $\alpha\in(0,1)$, writing
    $u:=\log(2p/\alpha)$, if $n$ is large enough, with
    probability at least $1-\alpha$,
    \begin{equation}\label{eq:nonasymptotic}
        \bigl\|\widehat{C}_n-C^{\star}\bigr\|_{\Fro}
        \ \leq\ c_0\,
        K\bigl(\varepsilon, p, \delta_{\max}\bigr)\,
        \sigma^{2}_{\max}\,
        \sqrt{\frac{pd\,(d+u)}{\gamma_{\min}n}} .
    \end{equation}
\end{theorem}

The sample size scales as $n^{-1/2}$, so the coupling is estimated at the same (parametric) rate as the marginals. The graph appears in the bound through $K (\varepsilon, p, \delta_{\max})$.
The rates inherit \Cref{prop:bound-diff} two-regime behavior.
When $\varepsilon>2\delta_{\max}\sigma^{2}_{\max}$ and the graph has bounded degree, the constant is bounded in $p$ and the bound is $\|\widehat{C}_n-C^{\star}\|_{\Fro}\lesssim d\sqrt{p/n}$, up to a factor $\{1+\log(p/\alpha)/d\}^{1/2}$. Otherwise it grows with $p$: at most as $p^{3/2}\varepsilon^{-1/2}$ for the complete graph, giving $p^{2}d/\sqrt{n\varepsilon}$.

The bound \eqref{eq:nonasymptotic} is one-sided and its constant is conservative. Studying the asymptotic behavior yields more precision: it allows us to compute the exact asymptotic variance and rate.

\begin{theorem}[Central limit theorem]\label{thm:clt}
    Let $\widehat{C}_n$ and $C^{\star}$ solve \eqref{eq:empirical_got}
    and \eqref{eq:entropic-ot-gaussian}, and assume
    $\sqrt{n}(\widehat{\margs}_n-\margs^{\star})\xrightarrow{d}N(0,\Gamma)$ on
    $\cX$. Then, with $\dphi:=\diffd\varphi_{\margs^{\star}}$,
    \begin{equation}\label{eq:clt}
        \sqrt{n}\bigl(\widehat{C}_n-C^{\star}\bigr)\ \xrightarrow{d}\
        Z\sim N\bigl(0,\dphi\,\Gamma\,\dphi^{\top}\bigr).
    \end{equation}
    If moreover $\Gamma\neq0$, then, with
    $T:=\tr(\dphi\Gamma\dphi^{\top})$,
    \begin{equation}\label{eq:rate}
        \bigl\|\widehat C_n-C^\star\bigr\|_{\Fro}=\Theta_{\PP}\bigl(\sqrt{T/n}\bigr),
        \qquad
        \tr\Gamma\ \leq\ T\ \leq\ \opnorm{\dphi}^{2}\,\tr\Gamma .
    \end{equation}
\end{theorem}
Statement \eqref{eq:clt} is the delta method ($\varphi$ is $\cC^{1}$ on $\cX^{++}$ by \Cref{thm:regularity-implicit}); the exact rate follows from the continuous mapping theorem together with $\sqrt{2/\pi}\sqrt{T}\leq\EE\|Z\|_{\Fro}\leq\sqrt{T}$.
The lower bound $T\geq\tr\Gamma$ uses $\dphi^{\top}\dphi\succeq\Id_{\cX}$, itself a consequence of $\diag\circ\dphi=\Id_{\cX}$, and shows in particular that $T>0$.
Under assumption \ref{assum:sampling} with $\gamma_i\equiv1$, $\Gamma$ has diagonal blocks $\Gamma_i(H)=2\Sigma_i^{\star}H\Sigma_i^{\star}$.
Hence $\tr\Gamma=\sum_i\{(\tr\Sigma_i^{\star})^{2}+\|\Sigma_i^{\star}\|_{\Fro}^{2}\} \asymp p\,d^{2}$ and \eqref{eq:rate} reads
\begin{equation*}
    \sigma^{2}_{\min}\,d\sqrt{p/n}
    \ \lesssim\ \bigl\|\widehat C_n-C^\star\bigr\|_{\Fro}
    \ \lesssim\ \opnorm{\dphi}\,\sigma^{2}_{\max}\,d\sqrt{p/n}
    \quad \text{ as } n \to \infty.
\end{equation*}
The upper bound is \eqref{eq:nonasymptotic} with the sharp constant
$\opnorm{\dphi}$, and matches the lower bound in the bounded-degree regime, where $\opnorm{\dphi}=O(1)$.

\subsection{Specification error}\label{subsec:specification}

Because $\chi$ is onto $\Pi^{++}(\margs^{\star})$ by \Cref{prop:identifiability}, the true covariance $C^{\rm true}$ is a weighted transport solution. Hence, there exists $\weights^{\rm true}$ such that
\begin{align*}
    C^{\rm true}=\chi(\weights^{\rm true}),
    \qquad \weights^{\rm true}_{ij}=-\varepsilon\bigl(C^{\rm true}\bigr)^{-1}_{ij}
    \quad (i\neq j).
\end{align*}
Solving \eqref{eq:entropic-ot-gaussian} amounts to choosing $\weights=\mathbf{1}_E\otimes I_d$, and the error corresponds to its distance to $\weights^{\rm true}$ in $\cW_d$. It follows that
\begin{equation}\label{eq:spec-error}
    \bigl\|C^{\star}-C^{\rm true}\bigr\|_{\Fro}
    \ \leq\ \min\Bigl\{
    \frac{\sqrt{2}\,\lambda_{+}^{2}}{\varepsilon}
    \bigl\|\weights^{\rm true}-\mathbf{1}_E\otimes I_d\bigr\|_{\cW_d}
    \ ,\
    B_{\varepsilon}(G,\margs^{\star})\Bigr\},
\end{equation}
valid whenever $C^{\rm true}$ is Markov with respect to $G$, i.e. whenever $\weights^{\rm true}$ is supported on $E$.

\begin{corollary}\label{cor:decomposition}
    Under assumption \ref{assum:sampling}, on the event of \Cref{thm:nonasymptotic} and
    with $u=\log(2p/\alpha)$,
    \begin{equation}\label{eq:decomposition}
        \bigl\|\widehat{C}_n-C^{\rm true}\bigr\|_{\Fro}
        \ \leq\
        \underbrace{c_0\,K(\varepsilon, p, \delta_{\max})
        \sigma^{2}_{\max}\sqrt{\tfrac{pd(d+u)}{\gamma_{\min}n}}}_{\text{estimation}}
        \ +\ \underbrace{
    \frac{\sqrt{2}\,\lambda_{+}^{2}}{\varepsilon}
    \bigl\|\weights^{\rm true}-\mathbf{1}_E\otimes I_d\bigr\|_{\cW_d} \wedge B_{\varepsilon}(G,\margs^{\star})}_{\text{specification}}.
    \end{equation}
\end{corollary}

The estimation error converges to zero when the number of samples increases, while the second term is nonzero unless $\weights^{\rm true}=\mathbf{1}_E\otimes I_d$, that is when the true coupling is the minimizer of the transport problem.

\subsection{Choice of the regularization parameter}\label{subsec:choice-epsilon}

By \Cref{prop:identifiability}, the regularization parameter $\varepsilon$ is the scale of the edges weights. Therefore, it is a modeling choice and cannot be uniquely determined from the graph and marginals only, unless some additional information on the joint distribution is given. 
 For instance, in the case of trajectories, $\varepsilon$ is the variance of the increments of the corresponding random walk and controls the roughness of its path. 
Specifically, all the points of the path $\varepsilon \in (0, +\infty) \mapsto C_\varepsilon^\star$ lying in $\cM(G, \margs)$ correspond to the same marginals and graph observations.
However, with additional information, $\varepsilon$ can be identified.
Hence, we propose a default choice for when no prior knowledge is available, and an exact choice for when information beyond the marginals is.

\textbf{Default choice.}
We make three observations.
\begin{itemize}[nosep]
    \item The regularization parameter $\varepsilon$ is homogeneous in $\margs$, so it has the unit of a variance.
    \item For the path graph with marginals $\sigma^2 I_d$, the stationary coupling $C_{ij}= \sigma^2 \rho^{|i-j|} I_d$ has $(C^{-1})_{i,i+1} = -\rho\,\sigma^{-2}(1-\rho^2)^{-1} I_d$ so it satisfies the KKT conditions for $\varepsilon = {\sigma}^2 \cdot (1 - \rho^2)/\rho$.
    \item For the complete graph with marginals $\sigma^2 I_d$, the coupling $C = \sigma^2 [(1-\rho)I_p + \rho \mathbf{1}_p\mathbf{1}_p^{\top} ]\otimes I_d$ has $(C^{-1})_{ij} = -\rho\,\sigma^{-2} ((1-\rho)(1+\delta\rho) )^{-1}I_d$ so it satisfies the KKT conditions for $\varepsilon = \sigma^2(1-\rho)(1+\delta\rho)/\rho$.
\end{itemize}
This suggests to use the following default regularization parameter: $\varepsilon = \bar{\sigma}^2 \cdot \bar{\delta} / 2$, where $\bar\sigma^{2}:=(pd)^{-1}\sum_i\tr\Sigma_i$ and $\bar\delta:=p^{-1}\sum_i\delta_i$. Inverting the closed-forms above shows that this choice induces a moderate coupling for the path and complete graphs.

\textbf{Choice of $\varepsilon$ with one cross-covariance.}
Information on the joint coupling beyond the marginals makes $\varepsilon$ identifiable. Specifically, the knowledge of one off-diagonal conditional covariance term determines it.
\begin{proposition}\label{prop:exact-epsilon}
    Let $\{i,j\} \in E$ and suppose that the covariance of $(X_i, X_j)$ conditionally on the remaining variables is known and has blocks $(\Sigma^{|}_{ii}, \Sigma^{|}_{ij}, \Sigma^{|}_{ji}, \Sigma^{|}_{jj})$. Then,
    \begin{align*}
        \varepsilon = \frac{\tr\bigl(\Sigma^{|}_{ii}
        -\Sigma^{|}_{ij}(\Sigma^{|}_{jj})^{-1}(\Sigma^{|}_{ij})^{\top}\bigr)}
        {\tr\bigl(\Sigma^{|}_{ij}(\Sigma^{|}_{jj})^{-1}\bigr)}.
    \end{align*}
    For $d=1$, it yields $\varepsilon = s_i s_j (1-r^2)/r$ where $s_i, s_j$ and $r$ are the standard deviations and correlation of $(X_i, X_j)$ conditionally on the remaining variables.
\end{proposition}
Therefore, if a joint covariance $\widehat C$ is available, $\varepsilon$ is chosen to be the average of precision entries $\varepsilon = -|E|d\bigl/\sum_{\{i,j\}\in E}\tr(\widehat C^{-1})_{ij}$.
If only unconditional cross-covariance entries are measured, $\varepsilon$ can still be chosen for the path and complete graphs with marginals $\sigma^2 I_d$ as shown above.

\section{Numerical study}\label{sec:numerical-study}

In this simulation study, we compare our theoretical findings with experimental results. Specifically, we focus on the effect of the input parameters of the problem -- the graph, the regularization constant $\varepsilon$, and the marginals -- on the joint covariance.
We also consider a statistical scenario where we only have empirical marginal covariances and explore the behavior of the corresponding estimated joint covariance \eqref{eq:empirical_got}. In particular, we study its convergence rates in the number of samples and its dependence on the ambient dimension.

To perform this study, we use the following tools and error measures.
Let $\margs^{\star} = (\Sigma_1^{\star}, \dots, \Sigma_p^{\star})$ be the marginal covariances with $p > 1$, $\varepsilon>0$ the entropic penalty parameter, and $G=(V,E)$ the graph.
As the graph and the entropy parameter influence how each variable relates to each other, it is of interest to assess their impact in terms of the nearest-neighbor correlation $\rho$ defined as $\rho_{i} = C_{i,i+1} / \sqrt{C_{i, i} C_{i+1, i+1}}$.
Furthermore, we showed in \Cref{sec:stability} that the stability of the solution is largely governed by the differential of the implicit function $\dphi$ and its operator norm \eqref{eq:diff-exact}. Therefore, we use it as a measure of the stability of the problem with respect to the different input parameters.
Finally, for the empirical scenario of Section \ref{sec:statistical-scenario}, we use the expected Frobenius norm to assess the convergence rate of the solution. Formally, if 
$\widehat{\margs}_n$ estimates $\margs^{\star}$, 
let $\widehat C_n$ and $C^{\star}$ solve the
empirical \eqref{eq:empirical_got} and population \eqref{eq:entropic-ot-gaussian} problems, we measure the error with the expectation $\EE \|\widehat{C}_n-C^{\star}\|_{\rm F}$.

\subsection{Stability of the problem: influence of the graph and of the entropic penalty}

To isolate the role of the graph $G$ and of the entropic penalty $\varepsilon$, we fix all remaining parameters of the problem: we set $p = 20$ and $\mu_1 = \ldots = \mu_p = N(0,1)$. The collection of graphs we study is the set of banded graphs introduced in Example \ref{ex:banded_graph}. In this model, the bandwidth encodes the connectivity of the graph. A recurring theme in this work is the stability of the solution depending on small variations of its margins. This stability can be read from the differential $\diffd \varphi$ of the implicit mapping. In this example, $\diffd \varphi$ is an element of $\mathcal{L}(\mathbb{R}^{20}, S_{20}(\mathbb{R}))$. Its operator norm $\|\diffd \varphi \|_{\op}$ quantifies the magnitude of this differential. In \Cref{prop:bound-diff}, we established a theoretical control of this quantity. In this section, exploiting a numerical solver and closed form \eqref{eq:diff-exact}, we compute $\|\diffd \varphi \|_{\op}$. 

\begin{figure}[h!]
    \centering
    \includegraphics[width=0.5\textwidth]{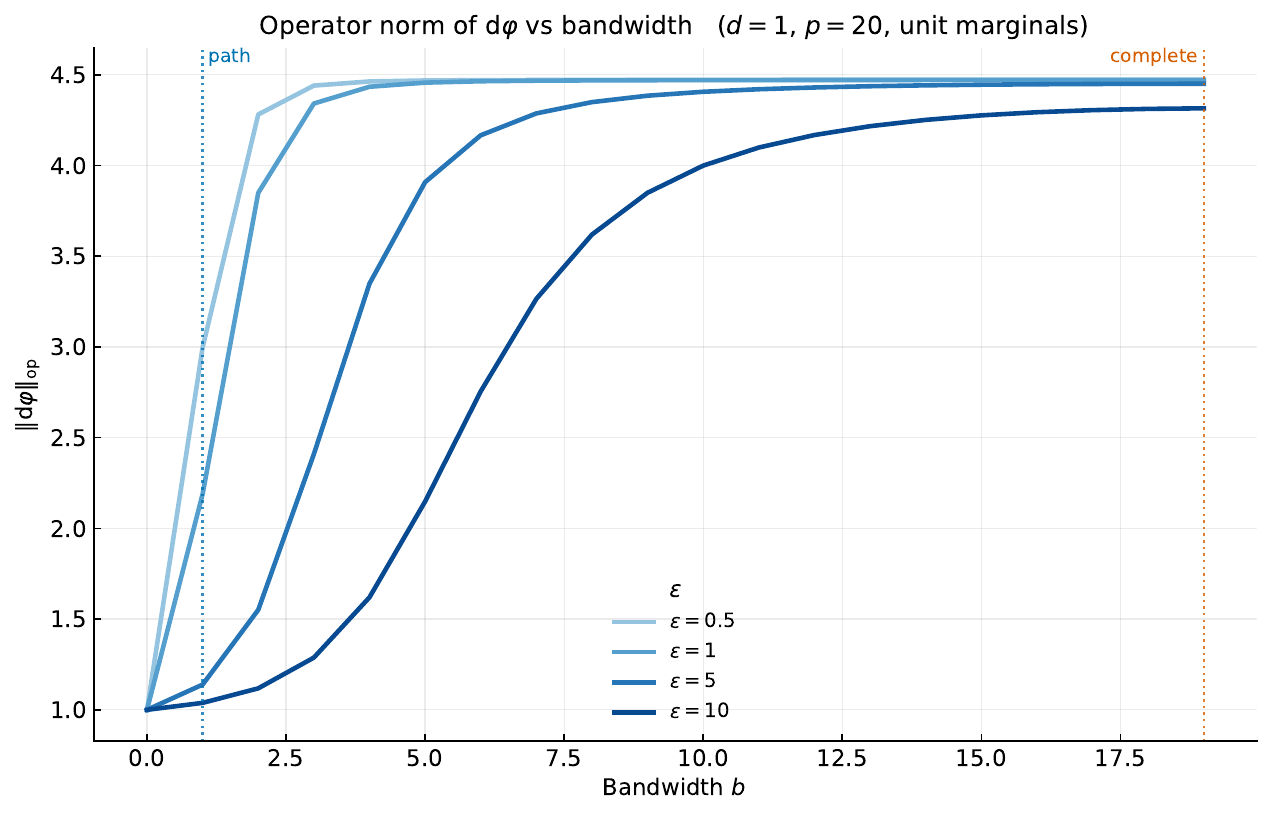}
    \caption{Operator norm of $\dphi$ with respect to the bandwidth of the banded graph.}
    \label{fig:opnorm_band}
\end{figure}

\begin{figure}[h!]
    \centering
    \begin{subfigure}{0.7\textwidth}
        \centering
        \includegraphics[width=\linewidth]{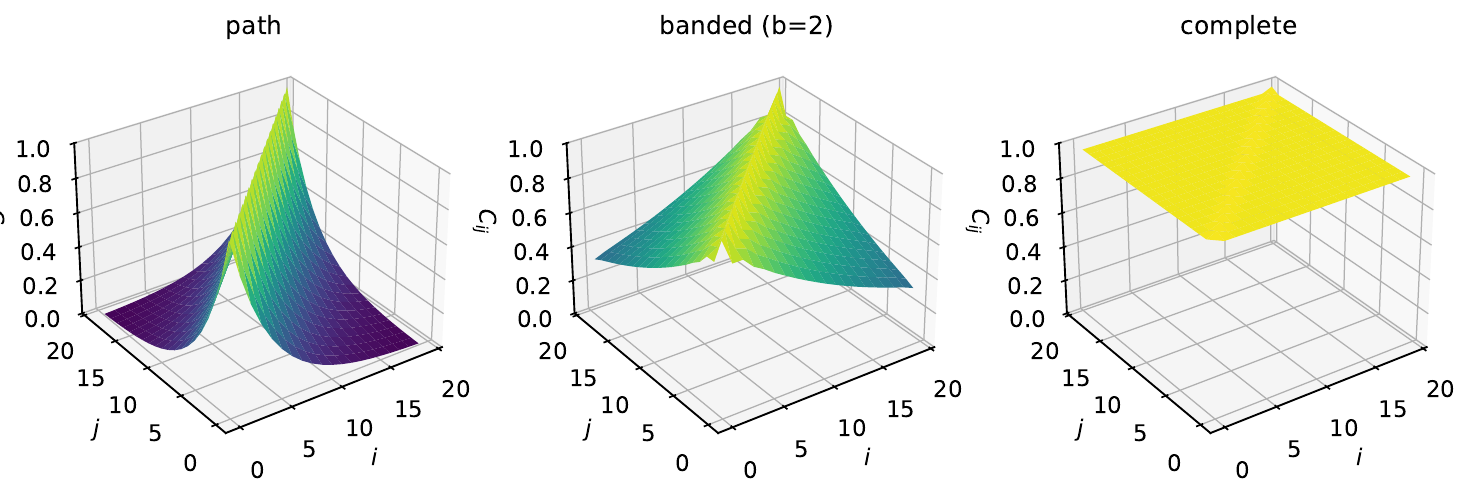}
    \end{subfigure}
    \begin{subfigure}{0.7\textwidth}
        \centering
        \includegraphics[width=0.97\linewidth]{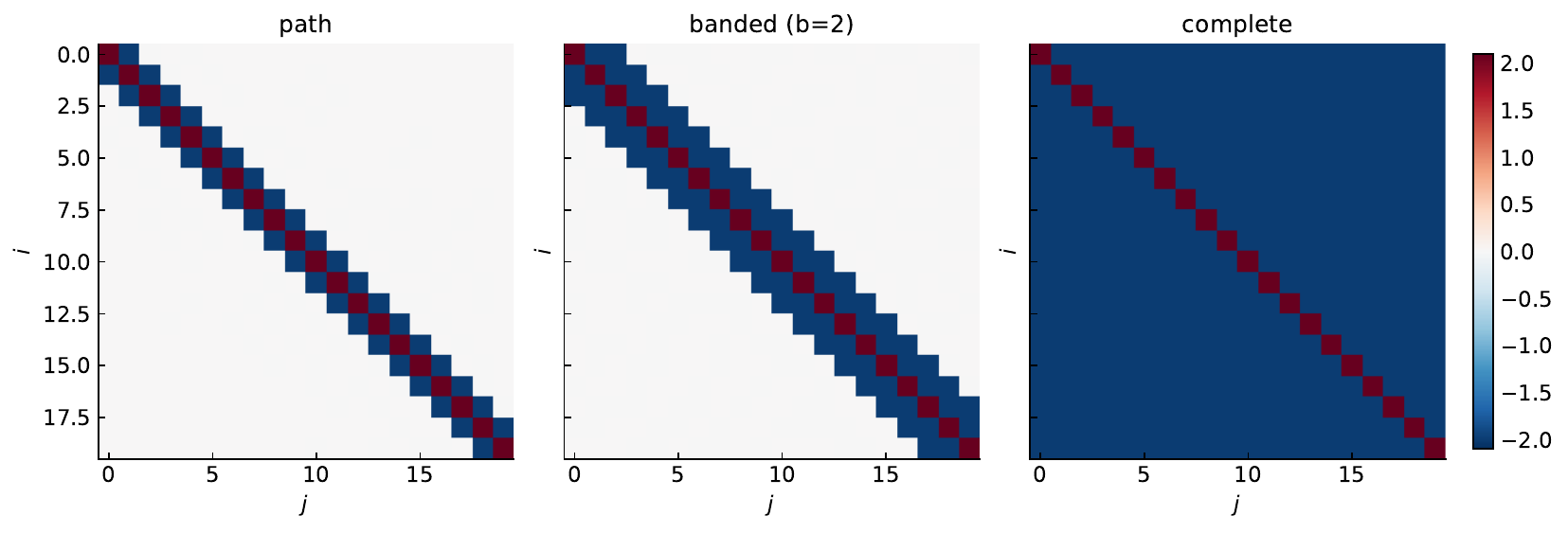}
    \end{subfigure}
    \begin{subfigure}{0.7\textwidth}
        \centering
        \includegraphics[width=\linewidth]{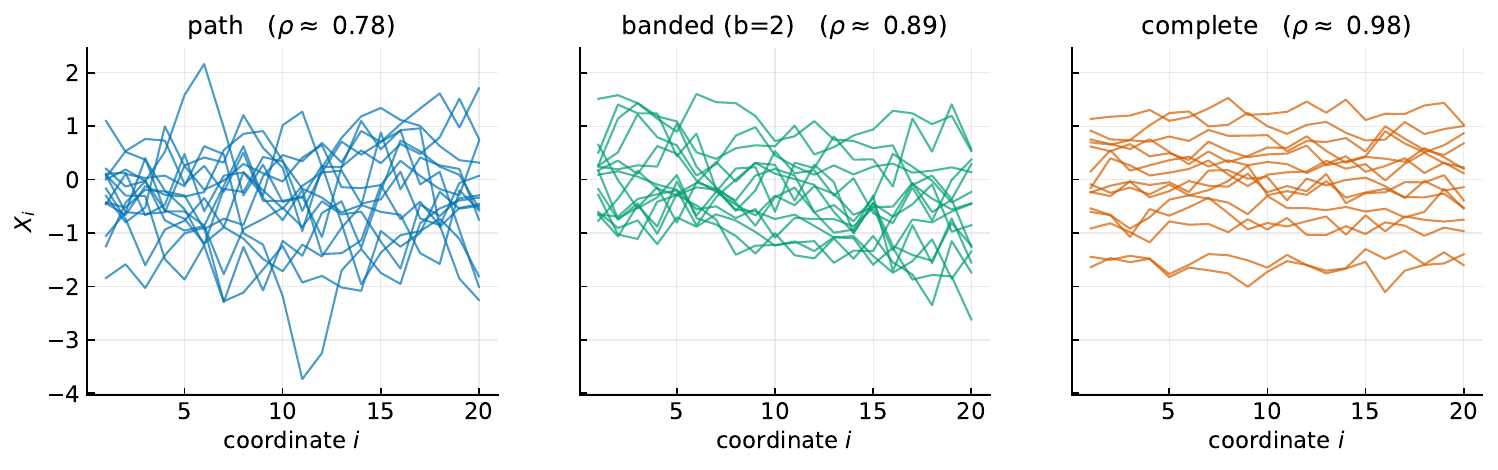}
    \end{subfigure}
    \caption{Solution of the multi-marginal problem for three graphs with $\varepsilon=0.5$.}
    \label{fig:solution_vs_graph}
\end{figure}

For $\varepsilon \in \{0.5,1,5, 10\}$, \Cref{fig:opnorm_band} displays the evolution of $\|\diffd \varphi \|_{\op}$ with band $b$ ranging between $0$ and $20$. It illustrates that increasing the connectivity of the graph increases the operator norm. Figure \ref{fig:solution_vs_graph} displays the impact of the graph on the reconstructed multi-coupling. Choosing a larger band of interaction increases correlation toward one. This increased correlation can also be read from the sample paths displayed in the last row of Figure \ref{fig:solution_vs_graph}. Sample paths corresponding to the complete graph feature smaller variations compared to sample paths reconstructed from the chain graph.

\begin{figure}[h!]
    \centering
    \includegraphics[width=0.5\textwidth]{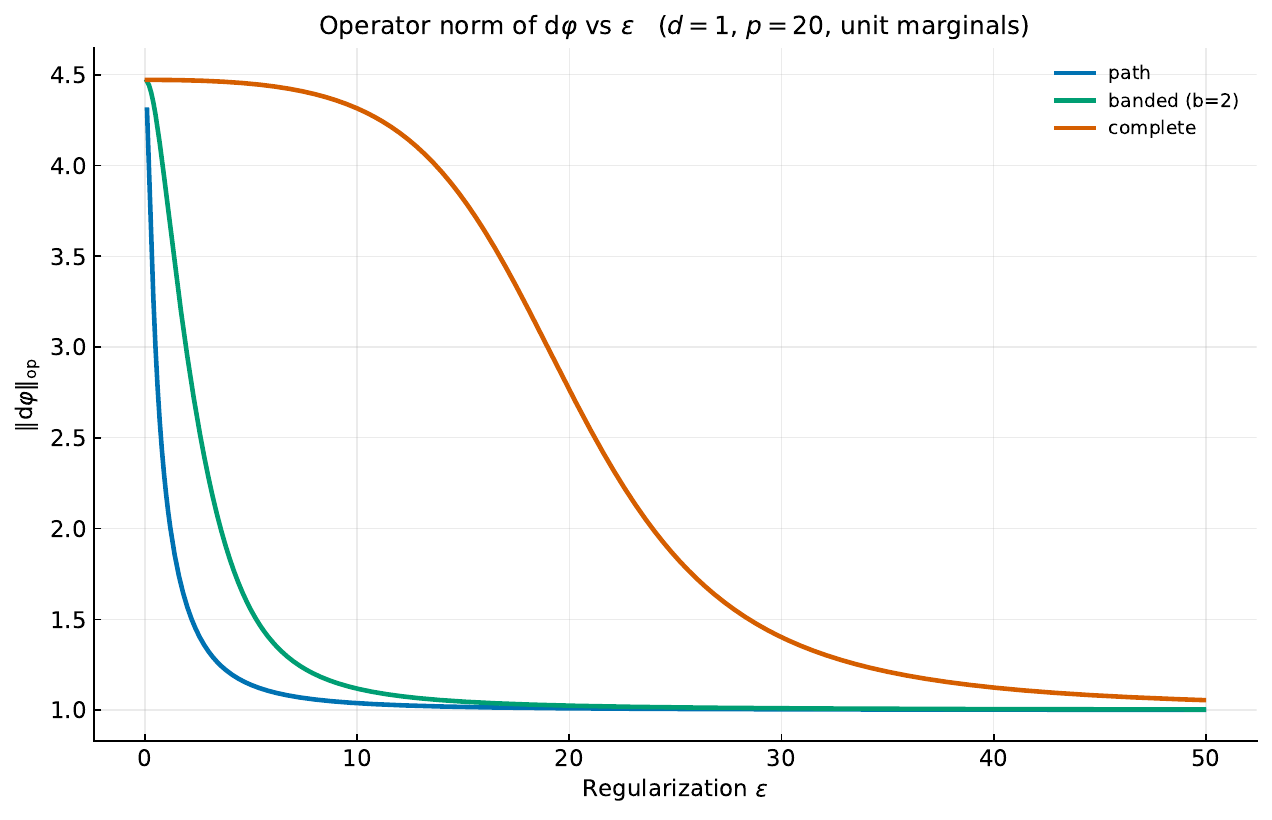}
    \caption{Operator norm of $\dphi$ with respect to the entropic penalty coefficient $\varepsilon$.}
    \label{fig:opnorm_epsilon}
\end{figure}

Another parameter that appears in the upper bound \eqref{eq:bound-diff} is the regularization parameter $\varepsilon > 0$. For three different banded graphs, we numerically compute $\| \diffd \varphi\|_{\op}$ for $\varepsilon \in (0, 50]$. The evolution of this operator norm as a function of $\varepsilon$ is displayed in Figure \ref{fig:opnorm_epsilon}. In this case, increasing $\varepsilon$ decreases the operator norm; or equivalently, increasing $\varepsilon$ stabilizes the problem.

\begin{figure}[h!]
    \centering
    \begin{subfigure}{0.7\textwidth}
        \centering
        \includegraphics[width=\linewidth]{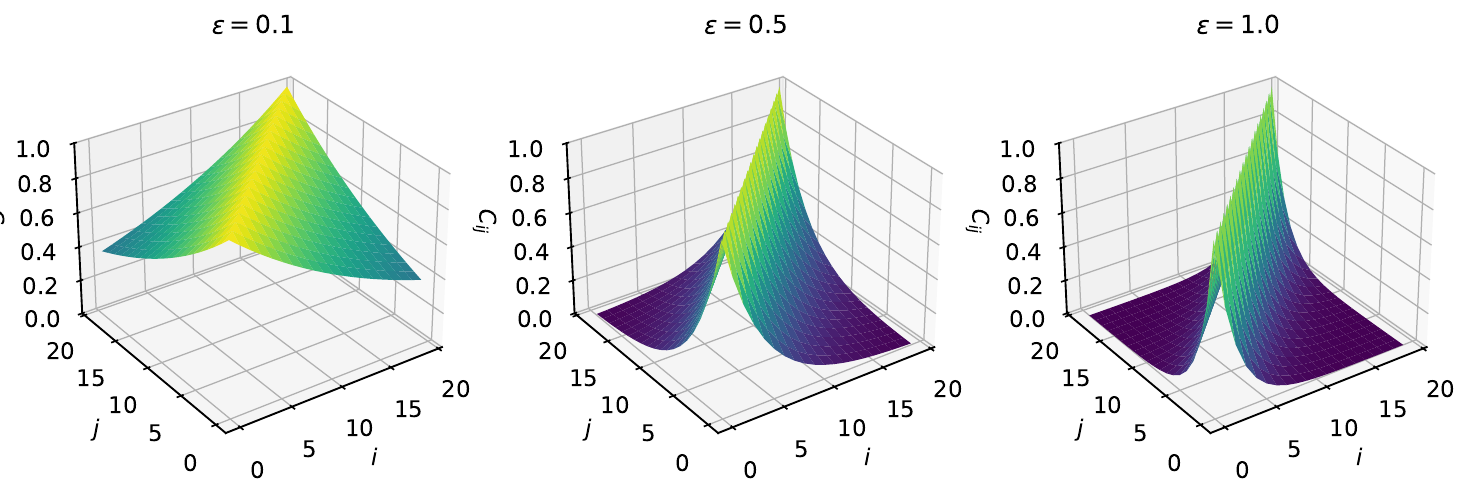}
    \end{subfigure}
    \begin{subfigure}{0.7\textwidth}
        \centering
        \includegraphics[width=\linewidth]{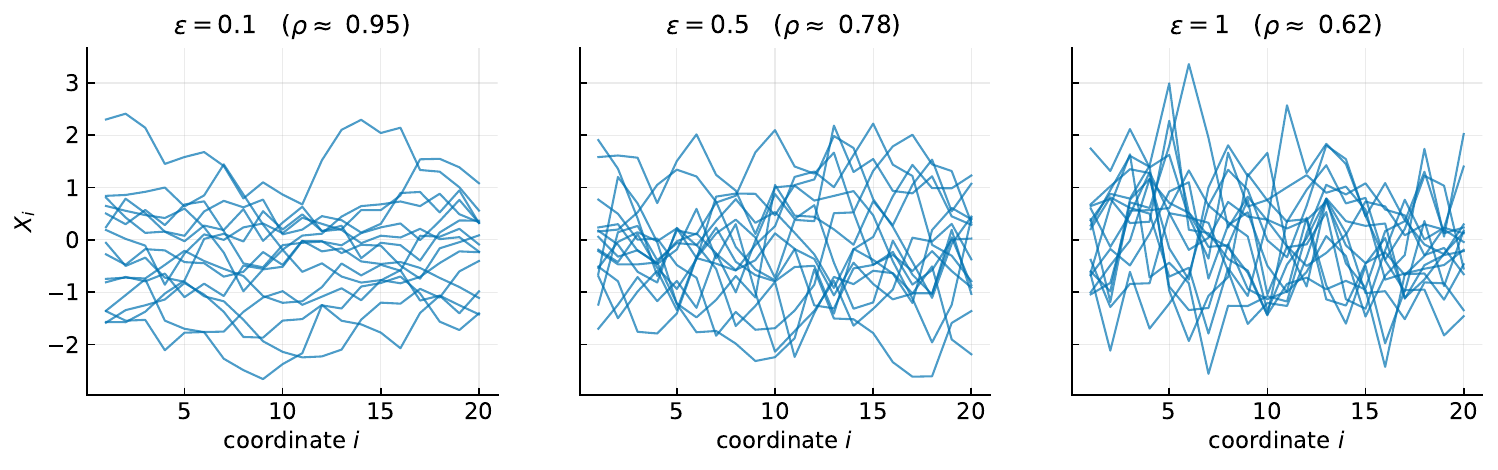}
    \end{subfigure}
    \caption{Evolution of the solution as $\varepsilon$ increases.}
    \label{fig:solution_vs_epsilon}
\end{figure}

Figure \ref{fig:solution_vs_epsilon} illustrates the impact of $\varepsilon$ on the  coupling. This figure shows the solution covariances as well as sample paths for increasing $\varepsilon$.  We end this section by comparing Figures \ref{fig:opnorm_band} and \ref{fig:opnorm_epsilon} with the theoretical result of Proposition \ref{prop:bound-diff}. The upper-bound \eqref{eq:bound-diff} captures that low connectivity of the graph and large $\varepsilon$ make the problem stable. However, our numerical computations suggest that our theoretical bounds could probably be improved in some cases. For example, Figure \ref{fig:opnorm_epsilon} shows that $\|\diffd \varphi\|_{\op}$ does not explode as $\varepsilon$ goes to zero.

\subsection{Empirical convergence and impact of the dimension}

We now study the empirical performance of the estimator introduced in \Cref{sec:statistical-scenario}. Recall that we have empirical marginal covariances $\widehat{\Sigma}_1,\ldots,\widehat{\Sigma}_p$, and that the estimated coupling $\widehat{C}_n$ is obtained by solving the empirical multi-marginal problem~\eqref{eq:empirical_got}. 
We compare the empirical convergence of $\widehat{C}_n$ towards the population solution $C^{\star}$ with the theoretical guarantees established in \Cref{sec:statistical-scenario}.

\begin{figure}[h!]
    \centering
    \includegraphics[width=0.5\textwidth]{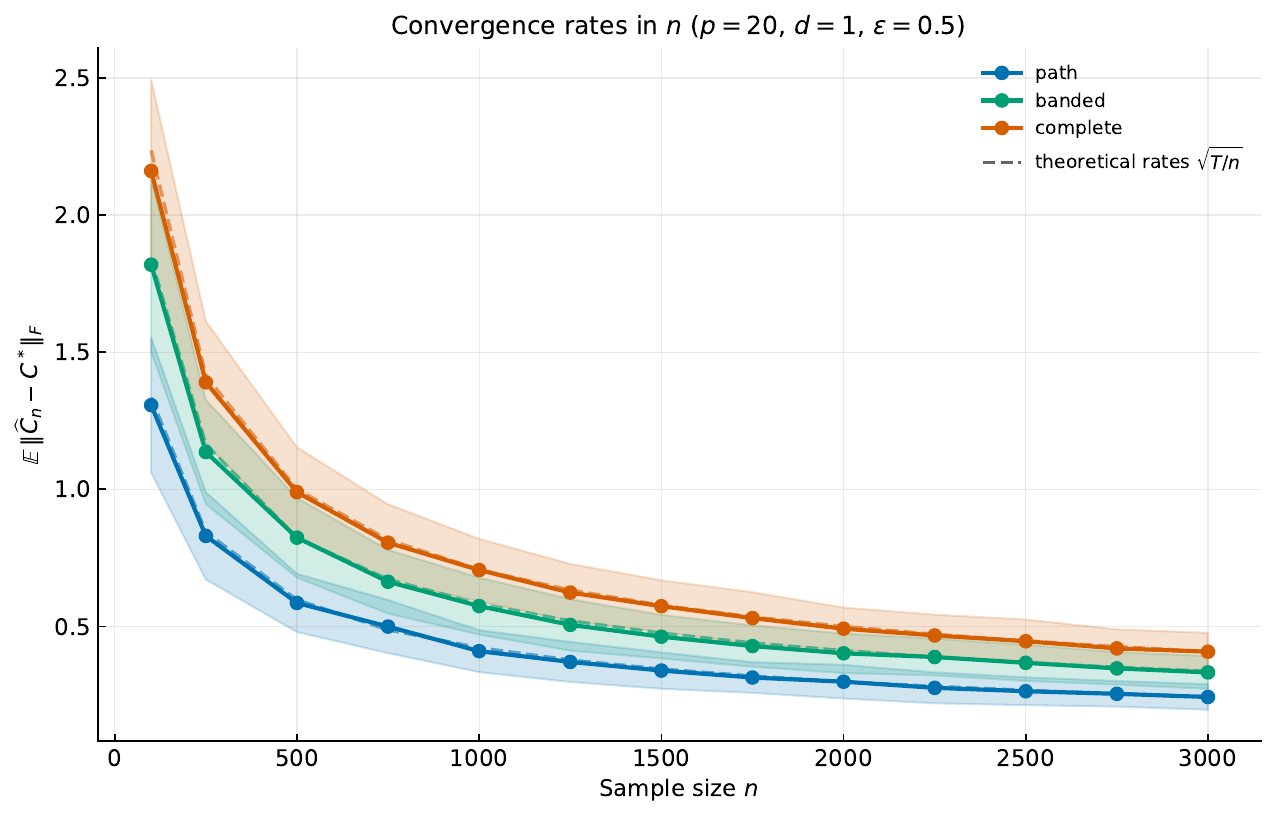}
    \caption{Convergence of the covariance estimate to the true covariance in expected Frobenius norm.}
    \label{fig:rates_n}
\end{figure}
Figure~\ref{fig:rates_n} displays the expected Frobenius error $\mathbb{E}\|\widehat{C}_n-C^{\star}\|_{\mathrm{F}}$ as a function of the sample size $n$. Across all graph topologies, the error decreases at the parametric rate $n^{-1/2}$ as proved in \Cref{thm:nonasymptotic}. However, the convergence constant depends on the graph topology: denser graphs exhibit larger estimation as in \Cref{prop:bound-diff}.

\begin{figure}[h!]
    \centering
    \includegraphics[width=0.9\textwidth]{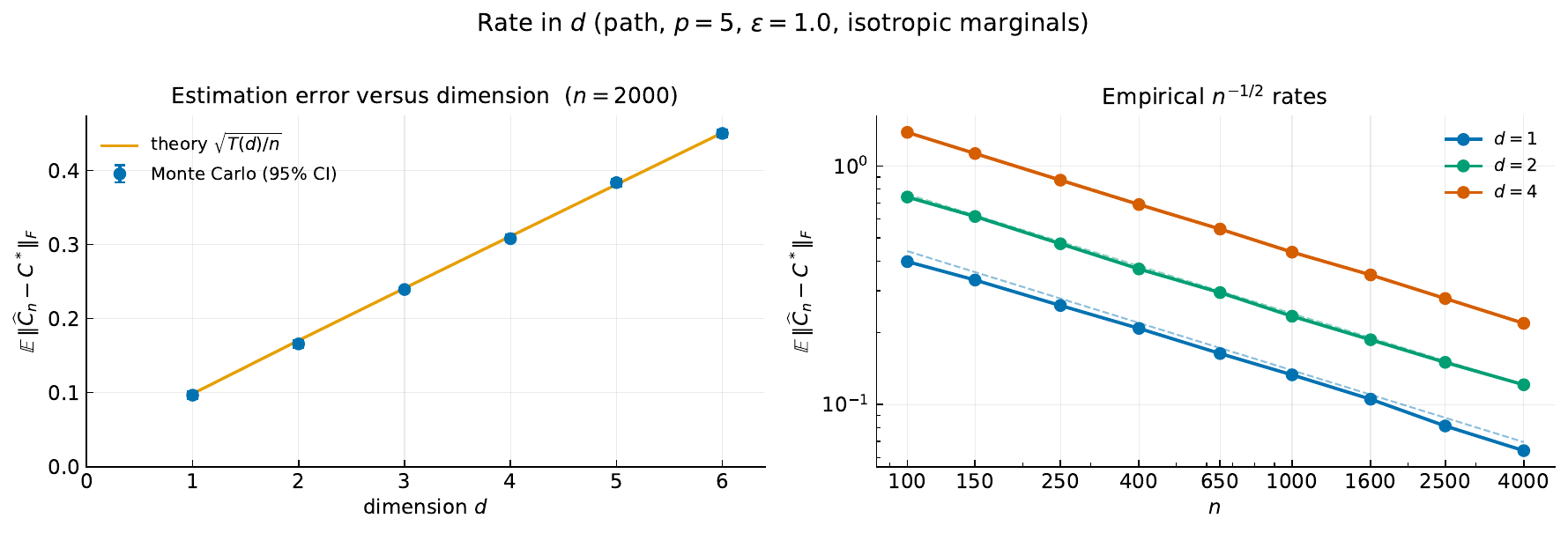}
    \caption{Dependence of the convergence rates in the ambient dimension $d$.}
    \label{fig:rates_d}
\end{figure}
Figure~\ref{fig:rates_d} illustrates the dependence of the estimation error on the ambient dimension $d$. The observed error grows linearly with $d$, matching the theoretical scaling \Cref{thm:nonasymptotic}. This linear dependence suggests that our approach could be extended to encompass infinite-dimensional Gaussian measures.

\newpage

\bibliographystyle{abbrvnat} 
\bibliography{Graph_OT}

\appendix

\section{Auxiliary material}

\subsection{Reminders on multi-marginal problems}\label{apx:subsec:multimargin-ot}

Optimal transport problems can be stated with various levels of generality. In this work, we aim for precise results in the concrete framework of Euclidean space. On $\mathbb{R}^d$, we consider $p$ probability measures $\mu_1,\ldots, \mu_p$ with $p\geq 2$ an integer. Let us also introduce a ground cost function\footnote{In the two marginals case, a ground cost $c(x,y)$ encodes the cost of moving one particle from $x$ to $y$. Following this interpretation in the multi-marginal problem, $c(x_1,\ldots,x_p)$ encodes the cost of one particle following trajectory $(x_1,\ldots,x_p)$.} $c : \mathbb{R}^{d\times p} \rightarrow \mathbb{R}_+$ that maps every element $\underline{x} = (x_1,\ldots, x_p) \in \mathbb{R}^{d\times p}$ to $c(x_1,\ldots, x_p) \in \mathbb{R}_+$. 
Denoting by $\proj_j$ the projection on the $j$th coordinate, i.e., $\proj_j(x_1,\ldots, x_p) = x_j$; the constraint set of the multi-marginal problem is 
\begin{equation}\label{eq:constraint_multimarge_measure_bis}
	\Pi(\underline{\mu}) := \{\pi \in \mathcal{P}(\mathbb{R}^{d\times p})~|~ \forall j \in \{1,\ldots, p\},~ \proj_j \# \pi = \mu_j \}.
\end{equation}
With these notations, the multi-marginal optimal transport problem with cost $c$ reads
\begin{equation}\label{eq:multimarge_ot_generic}
	W_c(\underline{\mu}) := \min_{\pi \in \Pi(\underline{\mu})} \int_{\mathbb{R}^{d\times p}} c(x_1,\ldots, x_p) \pi(\diffd x_1 \cdots \diffd x_p).
\end{equation}
Choosing $c_2(x_1,\ldots,x_p) = \sum_{i=1}^p \sum_{j=i+1}^p \|x_i - x_j\|^2$ yields the most famous multi-marginal problem, studied in \cite{gangbo1998optimal}, and given by
\begin{equation}\label{eq:multimarge_ot_euclidean}
W_2^2(\underline{\mu}) =  \min_{\pi \in \Pi(\underline{\mu})}  \sum_{i=1}^{p-1} \sum_{j=i+1}^p \int_{\mathbb{R}^{d\times p}} \|x_i-x_j\|^2 \pi(\diffd x_1 \cdots \diffd x_p).
\end{equation}
Both optimal transport problems \eqref{eq:multimarge_ot_generic} and \eqref{eq:multimarge_ot_euclidean} are written in the language of measure theory. They also have the following probabilistic formulations:
\begin{equation}
	W_c(\underline{\mu}) = \min_{\substack{(X_1,\ldots,X_p) \\ X_i \sim \mu_i}}  \mathbb{E}\left[ c(X_1,\ldots X_p)\right] \quad \text{and} \quad W_2^2(\underline{\mu}) = \min_{\substack{(X_1,\ldots,X_p) \\ X_i \sim \mu_i}} \sum_{i=1}^{p-1} \sum_{j=i+1}^p \mathbb{E}[\|X_i-X_j\|^2].
\end{equation}

In this document, we focus on transportation problems defined through ground costs related to the squared Euclidean distance.

\subsection{Reminders on graphical models and graph matrices}\label{app-sec:reminders_graph}

\paragraph{Gaussian graphical models}

We say that $G = (V, E)$ is the conditional independence graph of $X=(X_1, \dots, X_p) \in \RR^p$ if 
$V = \{1, \dots, p\}$ and 
\begin{equation*}
    \{ i,j \} \in E \iff X_i \notind 
    X_j \mid X_{-ij},
\end{equation*}
where $X_{-ij} = (X_k : k \in \{1,\ldots,p\} \setminus \{i,j\} )$.   
It is well known that, in the Gaussian case, the non-zero entries of the inverse covariance matrix give the edges in the graph. Formally, if $X \sim N(m, C)$ and $C$ has full rank, then 
\begin{equation*}
    X_i \ind X_j ~|~ X_{-ij}\quad \Longleftrightarrow \quad (C^{-1})_{ij} = 0.
\end{equation*}
\paragraph{Graph matrices}
The following definition is standard and can be found for example in \cite[Chap.~7]{nica2018brief}. 
\begin{definition}
	Given a graph $G = (\{1,\ldots,p\},E)$, the adjacency matrix $A(G) \in S_p(\mathbb{R})$ has each coefficient $(i,j)$ defined by 
	\begin{equation}\label{eq:adjacency_mat}
		A(G)_{ij} = {\bf 1}\{ \{i,j\}\in E \}.
	\end{equation}
	Denoting by $\delta_i := \sum_{j=1}^p {\bf 1}\{ \{i,j\}\in E \}$ the degree of vertex $i$, and $D(G)=\diag^*((\delta_i)_{1 \leq i \leq p})$ the diagonal matrix storing the degree of all vertices, the Laplacian matrix $L(G) \in S_p(\mathbb{R})$ is defined by 
	\begin{equation}\label{eq:laplacian_mat}
		L(G) = D(G) - A(G).
	\end{equation}
\end{definition}

\subsection{Dual problems for Gaussian multi-marginal optimal transport}\label{sec:dual_problem}

\begin{proposition}[Dual problem]\label{prop:dual_problem}
	Let $\mu_1,\ldots,\mu_p$ be $p$ centered Gaussian measures with respective covariances matrices $\Sigma_1,\ldots, \Sigma_p$. When $\varepsilon > 0$, exploiting  matrix reduction \eqref{eq:entropic-ot-gaussian}, the graphical optimal transport problem \eqref{eq:entropy_graph_transport} has dual problem
\begin{equation}\label{eq:dual_prob_laplace}
	W_{G}^{\varepsilon}(\underline{\Sigma}) = \sup_{\Lambda_1,\ldots,\Lambda_p} \sum_{i=1}^p \langle - \Sigma_i, \Lambda_i \rangle_{\rm F} + \varepsilon \log \det (L + \diag^*(\underline{\Lambda})) + m_{\mathcal{D}}(\varepsilon),
\end{equation}
where $m_{\mathcal{D}}(\varepsilon)=\varepsilon \sum_{i=1}^p \log \det(\varepsilon^{-1}\Sigma_i)$. In the un-regularized case, problem \eqref{eq:unregularized-ot-gaussian} has dual counterpart
\begin{equation}\label{eq:dual_noreg}
W_{G}(\underline{\Sigma})	=  \sup_{\Lambda_1,\ldots,\Lambda_p} \sum_{j = 1}^p \langle -\Sigma_i, \Lambda_i \rangle_{\rm F} \quad \text{such that $L + \diag^*(\underline{\Lambda})$ is positive semidefinite.}
\end{equation}
\end{proposition}

\begin{proof}
    We introduce $\mathcal{F} : S_p(M_d(\mathbb{R})) \rightarrow \mathbb{R} \cup\{+\infty\} $ the objective function defined by 
    \begin{equation}\label{eq:objective_primal}
    \mathcal{F}(C) := \tr(LC) -\varepsilon \log \det(C).
    \end{equation}
    With $\mathcal{F}$ now introduced, we can rewrite the entropic problem as follows.
    \begin{equation}\label{eq:primal_objective_and_const}
    W_{G,\varepsilon}(\underline{\mu}) = \min_{C \in \Pi(\underline{\Sigma})}  \mathcal{F}(C) + m_{\mathcal{P}}(\varepsilon) \quad \text{where} \quad  m_{\mathcal{P}}(\varepsilon):= \varepsilon\sum_{i=1}^p\log \det (\Sigma_i)
    \end{equation}
    is a constant independent of the covariance $C$.
	Our starting point is formulation \eqref{eq:primal_objective_and_const}, without the additive constant $m_{\mathcal{P}}(\varepsilon)$. This is a convex problem with equality constraints. Thus, the Lagrangian function $\mathcal{L}$ associated to this problem is defined at primal variable $C \in S_p(M_d(\mathbb{R}))$ and multipliers $\underline{\Lambda}\in S_d(\mathbb{R})^p$ by 
	\begin{align}\label{eq:lagrangian_function}
		\mathcal{L}(C,\underline{\Lambda})& = \mathcal{F}(C) + \sum_{i=1}^p \langle C_{ii}- \Sigma_i, \Lambda_i \rangle_{\rm F} \\
		& = \tr(LC) - \varepsilon \log \det(C) + \sum_{i=1}^p \langle C_{ii}- \Sigma_i, \Lambda_i \rangle_{\rm F} \nonumber \\
		& = \sum_{i=1}^p \langle -\Sigma_i, \Lambda_i\rangle_{\rm F} + \tr(\left[L+\diag^*(\underline{\Lambda})\right]C) - \varepsilon \log \det(C). \nonumber
	\end{align}
	The dual problem associated to the Lagrangian function \eqref{eq:lagrangian_function} is 
	\begin{equation}\label{eq:inproof_dualprob}
		\sup_{\underline{\Lambda} \in S_d(\mathbb{R})^p} \mathcal{D}(\underline{\Lambda}) \quad \text{where} \quad \mathcal{D}(\underline{\Lambda}) := \inf_{C \in S_p(M_d(\mathbb{R}))} \mathcal{L}(C,\underline{\Lambda})
	\end{equation}
	is called the dual function. We now derive an explicit expression for $\mathcal{D}$. Let us fix a multiplier $\underline{\Lambda}$ and observe that $\mathcal{L}(\cdot, \underline{\Lambda})$, as a function of the first variable only is convex. The problem of computing the dual function reduces to the search of critical points. Differentiating the Lagrangian with respect to $C$ yields
	\begin{equation*}
		\nabla_C\mathcal{L}(C, \underline{\Lambda}) = L + \diag^*(\underline{\Lambda}) - \varepsilon C^{-1}.
	\end{equation*}
	Hence, the optimality condition $\nabla_C\mathcal{L}(C_{\underline{\Lambda}}, \underline{\Lambda}) = 0$ is equivalent to 
	\begin{equation}\label{eq:relation_primal_dual}
		C_{\underline{\Lambda}}^{-1} = \frac{1}{\varepsilon}\left[ L + \diag^*(\underline{\Lambda}) \right],
	\end{equation}
	or alternatively
	\begin{equation*}
		C_{\underline{\Lambda}} = \varepsilon \left[L +\diag^*(\underline{\Lambda})\right]^{-1}.
	\end{equation*}
	From this, we write the dual function
	\begin{align*}
		\mathcal{D}(\underline{\Lambda}) & = \mathcal{L}(C_{\underline{\Lambda}}, \underline{\Lambda}) \\
		& = \sum_{i=1}^p \langle -\Sigma_i, \Lambda_i\rangle_{\rm F} + \varepsilon\tr(\left[L+\diag^*(\underline{\Lambda})\right]\left[L+\diag^*(\underline{\Lambda})\right]^{-1}) - \varepsilon \log \det( \varepsilon[L +\diag^*(\underline{\Lambda})]^{-1}) \\
		& = \sum_{i=1}^p  \langle -\Sigma_i, \Lambda_i\rangle_{\rm F} + \varepsilon \log \det(L +\diag^*(\underline{\Lambda})) + \varepsilon pd (1 - \log(\varepsilon)).
	\end{align*}
	We now remind the additive constant 
	\begin{equation*}
		m_{\mathcal{P}}(\varepsilon) = \varepsilon \left[\log \det(\diag^*(\underline{\Sigma}))-pd\right] = \varepsilon \sum_{i=1}^p \log \det(\Sigma_i) - \varepsilon pd
	\end{equation*}
	associated to the primal problem. We define the constant associated to the dual problem  $m_{\mathcal{D}}(\varepsilon)$ as  
	\begin{align*}
		m_{\mathcal{D}}(\varepsilon) &:= m_{\mathcal{P}}(\varepsilon)  + \varepsilon pd (1 - \log(\varepsilon)) \\
		& = \varepsilon \sum_{i=1}^p \left( \log \det(\Sigma_i) - d\log(\varepsilon) \right) \\
		& = \varepsilon \sum_{i=1}^p \log \det(\varepsilon^{-1}\Sigma_i).
	\end{align*}
	Collecting the pieces, the dual problem associated to \eqref{eq:entropic-ot-gaussian} is 
	\begin{equation*}
		W_{G}^{\varepsilon}(\underline{\Sigma}) = \sup_{\Lambda_1,\ldots,\Lambda_p} \sum_{i=1}^p \langle - \Sigma_i, \Lambda_i \rangle_{\rm F} + \varepsilon \log \det (L + \diag^*(\underline{\Lambda})) + m_{\mathcal{D}}(\varepsilon).
	\end{equation*}
When $\varepsilon = 0$, the problem has a different structure: we have a linear objective function, and the positivity constraint is not guaranteed by the $\log\det$ function, as it is in the regularized case. Additionally to the dual variables $\underline{\Lambda} \in S_d(\mathbb{R})^p$ for equality constraints, the primal variable $C $ should be restricted to the cone of positive semidefinite matrices $S_p^+(M_d(\mathbb{R}))$. Thus, at $(C,\underline{\Lambda}) \in S_p^+(M_d(\mathbb{R})) \times S_d(\mathbb{R})^p$, the Lagrangian reads  
\begin{align*}
	\mathcal{L}(C,\underline{\Lambda}) & = \langle L, C \rangle_{\rm F} + \langle \diag(C) - \underline{\Sigma}, \underline{\Lambda}\rangle  \\
	& = \sum_{i=1}^p \langle - \Sigma_i, \Lambda_i \rangle_{\rm F} + \langle L, C \rangle_{\rm F} + \langle C, \diag^*(\underline{\Lambda}) \rangle_{\rm F} \\
	& = \sum_{i=1}^p \langle - \Sigma_i, \Lambda_i \rangle_{\rm F} + \langle L + \diag^*(\underline{\Lambda}) \rangle_{\rm F}.
\end{align*}
The dual problem is 
\begin{align}
	\sup_{\underline{\Lambda} \in S_d(\mathbb{R})^p} \min_{C \in S_p^+(M_d(\mathbb{R}))}\mathcal{L}(C,\underline{\Lambda}).
\end{align}
And at $\underline{\Lambda} \in S_d(\mathbb{R})^p$,
\begin{align*}
\min_{C \in S_p^+(M_d(\mathbb{R}))}\mathcal{L}(C,\underline{\Lambda}) & = \sum_{i=1}^p \langle - \Sigma_i, \Lambda_i \rangle_{\rm F} + \min_{C \in S^+} \langle L + \diag^*(\underline{\Lambda}), C \rangle_{\rm F} \\
&  = \begin{cases}
	\sum_{i=1}^p \langle - \Sigma_i, \Lambda_i \rangle_{\rm F} & \text{if }  L + \diag^*(\underline{\Lambda}) \geq 0 \\
	- \infty & \text{otherwise}.
\end{cases}
\end{align*}
This last computation shows that we should restrict to dual variables such that $ L + \diag^*(\underline{\Lambda}) \geq 0$. This yields the dual problem 
\begin{equation*}
\sup_{\underline{\Lambda} \in S_d(\mathbb{R})^p}  \sum_{i=1}^p \langle  \Sigma_i, -\Lambda_i \rangle_{\rm F} \quad \text{s.t.} \quad L + \diag^*(\underline{\Lambda}) \geq 0.
\end{equation*}
\end{proof}

\section{Proofs of technical results}


\subsection{Omitted proofs from Section \ref{sec:multimarge_ot}}\label{subsec:proof_gauss_condindep}

\begin{proof}[Proof of \Cref{prop:unregularized-ot-gaussian}]
	First, we introduce $I_G : \Pi(\underline{\mu}) \rightarrow \mathbb{R}$ the objective function associated with $W_G(\underline{\mu})$. It is defined at $\pi \in \Pi(\underline{\mu})$ by 
	\begin{equation*}
		I_G(\pi) := \sum_{\{i,j\} \in E} \int_{\mathbb{R}^{d \times 2}} \|x_i - x_j\|^2 \pi(\diffd x_i \diffd x_j).
	\end{equation*}
	If $\{i,j\} \in E$, we write $i \sim j$. With this notation, we can rewrite the objective function at $\pi$ as
	\begin{equation*}
		I_G(\pi) =\frac{1}{2}  \sum_{i=1}^p \sum_{j \sim i} \int_{\mathbb{R}^{d \times 2}} \|x_i- x_j\|^2 \pi(\diffd x_i \diffd x_j).
	\end{equation*}
	Expanding the squared Euclidean cost yields
	\begin{align*}
		I_G(\pi) &= \frac{1}{2} \sum_{i=1}^p \sum_{j \sim i} \left( \int_{\mathbb{R}^{d}} \|x_i\|^2 \mu_i(\diffd x_i) + \int_{\mathbb{R}^{d}} \|x_j\|^2 \mu_j(\diffd x_j) - 2 \int_{\mathbb{R}^{d \times 2}} \langle x_i, x_j \rangle \pi(\diffd x_i \diffd x_j)\right).
	\end{align*}
	Noticing that for every index $i \in \{1,\ldots, p\}$ the equality
	\begin{equation*}
		\int_{\mathbb{R}^{d}} \|x_i\|^2 \mu_i(\diffd x_i) = \tr (\Sigma_i)
	\end{equation*}
	holds true,	we can rewrite 
	\begin{equation*}
		\sum_{i=1}^p \sum_{j \sim i} \int_{\mathbb{R}^{d}} \|x_i\|^2 \mu_i(\diffd x_i)  = 	\sum_{i=1}^p \sum_{j \sim i} \tr (\Sigma_i) = \sum_{i=1}^p \delta_i \tr (\Sigma_i).
	\end{equation*}
	As the graph is undirected, we have 
	\begin{equation*}
		\sum_{i=1}^p \sum_{j\sim i}   \int_{\mathbb{R}^{d}} \|x_j\|^2 \mu_j(\diffd x_j) = \sum_{j=1}^p \sum_{i \sim j} \tr(\Sigma_j)  =  \sum_{j=1}^p \delta_j \tr ( \Sigma_j ).
	\end{equation*}
	Collecting the pieces, we can write the objective function 
	\begin{equation}\label{eq:degree_correlation_obj}
		I_G(\pi)= \sum_{i =1}^p \delta_i \tr(\Sigma_i) - \sum_{i=1}^p \sum_{j \sim i}\int_{\mathbb{R}^{d \times 2}} \langle x_i, x_j \rangle \pi(\diffd x_i \diffd x_j).
	\end{equation}
	In last formula \eqref{eq:degree_correlation_obj}, the first term is independent of the multicoupling $\pi$. We now focus on the second term involving the inner product.
	for $\pi \in \Pi(\mu_1,\ldots, \mu_p)$, we denote by $C_\pi$ its covariance matrix defined by 
\begin{equation*}
	C_\pi := \mathbb{E}_{X\sim \pi}[(X_1,\ldots,X_p) \otimes (X_1,\ldots,X_p)].
\end{equation*}

We point out that $C_\pi$ is a $p \times p$ block matrix, with each block $(C_\pi)_{i,j}$ explicitly defined by 
\begin{equation*}
	(C_\pi)_{i,j} = \mathbb{E}[X_i \otimes X_j] \quad \text{where} \quad (X_1,\ldots, X_p) \sim\pi.
\end{equation*}
 With this covariance $C_\pi$ introduced, we can write 
\begin{align}
	2\sum_{\{i,j\} \in E}  \int_{\mathbb{R}^{d \times 2}} \langle x_i, x_j \rangle \pi(\diffd x_i \diffd x_j) & =  2 \sum_{\{i,j\}\in G} \tr(\mathbb{E}[X_i \otimes X_j]) \nonumber \\
	& = 2 \sum_{\{i,j\} \in G} \tr( C(\pi)_{i,j}) \nonumber \\
	& = \langle A(G) \otimes I_d, C(\pi) \rangle_{\rm F},
\end{align}
where $A(G)$ is the adjacency matrix of $G$, $I_d$ is the identity operator on $E$, and $\langle \cdot, \cdot \rangle_{\rm F}$ is the Frobenius inner product defined at $A,B$ by $\langle A, B \rangle_{\rm F} := \tr(AB^T)$. Regarding the first term in \eqref{eq:graph_ot_correlation_max}, with a similar computation we derive
\begin{align*}
	\sum_{i=1}^p \delta_i \mathbb{E} \left[\|X\|^2\right]& = \sum_{i=1}^p \delta_i \tr \mathbb{E}[X_i \otimes X_i] \\
	& = \sum_{i=1}^p\delta_i \tr(C_\pi)_{ii} \\
	& = \langle D(G)\otimes I_d, C_\pi \rangle_{\rm F},
\end{align*}
with $D(G)$ the degree matrix. Collecting the pieces, we can rewrite the objective function associated to the graphical optimal transport model as
\begin{equation}
	I_G(\pi) = \langle D(G)\otimes I_d - A(G)\otimes I_d, C_\pi \rangle_{\rm F} = \langle L(G)\otimes I_d, C_\pi \rangle_{\rm F}.
\end{equation}
As $\pi$ has marginals $\mu_1, \ldots, \mu_p$, and each $\mu_i$ has covariance $\Sigma_i$; for every diagonal block, the equality $C(\pi)_{ii} = \Sigma_{i}$ must hold true. That is $C(\pi)$ belongs to the set $\Pi(\underline{\Sigma})$. To conclude, notice that the objective function $I_G$ depends only of the covariance of the coupling. Furthermore, for any covariance matrix $C \in \Pi(\underline{\Sigma})$, the centered Gaussian measure $N(0,C)$ is an admissible coupling in $\Pi(\underline{\mu})$. Thus, instead of optimizing over the set of admissible couplings $\Pi(\underline{\mu})$, we can optimize over admissible covariances $\Pi(\underline{\Sigma})$, leading to formulation \eqref{eq:unregularized-ot-gaussian} of the graphical optimal transport problem.

\end{proof}

\medskip

\begin{proof}[Proof of \Cref{prop:entropic-ot-gaussian}]
At $\pi \in \Pi(\mu_1,\ldots, \mu_p)$, the objective function associated to problem \eqref{eq:entropy_graph_transport} is 
\begin{equation}
	I_G^{\varepsilon}(\pi) = \sum_{\{i,j\} \in E } \int_{\mathbb{R}^{d \times 2}} \|x_i-x_j\|^2 \pi(\diffd x_1 \cdots \diffd x_p) + 2\varepsilon \KL(\pi |N(\diag(\underline{\Sigma}))).
\end{equation}
The first term in the previous expression has already been studied in Proposition \ref{prop:unregularized-ot-gaussian}, where we derived the equality
\begin{equation*}
 \sum_{\{i,j\} \in E}	\int_{\mathbb{R}^{d \times 2}} \|x_i - x_j\|^2 \pi(\diffd x_1\cdots \diffd x_p) = \langle L, C(\pi) \rangle_{\rm F}
\end{equation*} 
with $C(\pi)$ is the covariance matrix of $\pi$. Moreover, we have shown that we can reduce the search space $\Pi(\mu_1,\ldots, \mu_p)$ to centered Gaussian measures with covariance matrix in $\Pi(\underline{\Sigma})$. Now, we exploit that for Gaussian measures the Kullback-Leibler divergence has closed-form 
\begin{equation}
	\KL(N(0,C)|N(0,\diag^*(\underline{\Sigma}))) =  \frac{1}{2} \left(\tr(\diag^*(\underline{\Sigma}^{-1})C - \Id) - \log \det(\diag^*(\underline{\Sigma}^{-1})C) \right).
\end{equation} 
And as $C \in \Pi(\Sigma_1,\ldots, \Sigma_p)$, we have the equality $\tr(\diag^*(\underline{\Sigma}^{-1}C)) = \tr(\Id)$. Thus, we can rewrite the regularizing term
\begin{equation*}
	2 \varepsilon \KL(N(0, C)|N(0, \diag^*(\underline{\Sigma}))) = -\varepsilon \log \det(\diag^*(\underline{\Sigma}^{-1})C) .
\end{equation*}

Collecting the pieces, we deduce the announced reduction \eqref{eq:entropic-ot-gaussian}.
\end{proof}

\medskip

\begin{proof}[Proof of \Cref{eq:dependence-budget}]
    Write $C^\star = C^\star_{\varepsilon}$. Let $C \in \Pi(\margs) \setminus \{C^{\star}\}$ such that $\cD(C) \leq \cD(C^\star)$. By definition of $C^\star$, and because it is unique, $\langle L, C^\star \rangle + 2\varepsilon \cD(C^\star) < \langle L, C \rangle + 2\varepsilon \cD(C)$.
    Hence,
    \begin{align*}
        \langle L, C^\star \rangle < \langle L, C \rangle + 2\varepsilon\{\cD(C) - \cD(C^\star)\} \leq \langle L, C \rangle.
    \end{align*}
    As $C^\star$ is feasible for the problem it follows that it is its unique solution.
\end{proof}

\medskip

\begin{proof}[Proof of \Cref{thm:graph_transport_optimality_condition}]
	As the primal problem is convex, strong duality holds. That is
	\begin{equation*}
 \min_{C \in \Pi(\underline{\Sigma})}  \mathcal{F}(C) = \max_{\underline{\Lambda}\in S_d(\mathbb{R})^p}\mathcal{D}(\underline{\Lambda}).
	\end{equation*}
And from the proof of Proposition \ref{prop:dual_problem}, we know that
\begin{equation}
	\max_{\underline{\Lambda}\in S_d(\mathbb{R})^p}\mathcal{D}(\underline{\Lambda})= \max_{\underline{\Lambda}\in S_d(\mathbb{R})^p} \mathcal{L}(C_{\underline{\Lambda}}, \underline{\Lambda}) \quad \text{with} \quad C_{\underline{\Lambda}} = \varepsilon \left[L +\diag^*(\underline{\Lambda})\right]^{-1}.
\end{equation} 
To conclude, we need to establish that for optimal $\underline{\Lambda}$, the global covariance $C_{\underline{\Lambda}}$ satisfies primal feasibility. I.e., that for every diagonal blocks of $C_{\underline{\Lambda}}$, the equality $(C_{\underline{\Lambda}})_{ii} = \Sigma_i$ holds true.\\

Set $i \in \{1,\ldots, p\}$. At optimality $\nabla_{\underline{\Lambda}}\mathcal{D}(\underline{\Lambda}) = 0$. And in particular the gradient with respect to the $i$th variable $\nabla_{\Lambda_i}\mathcal{D}(\underline{\Lambda})=0$. Let us compute $\nabla_{\Lambda_i}\mathcal{D}(\underline{\Lambda})$ at an arbitrary $\underline{\Lambda}$.
Remind that $\mathcal{D}(\underline{\Lambda}) = \sum_{j=1}^p \langle -\Sigma_j, \Lambda_j \rangle_{\rm F} + \varepsilon \log \det (L+\diag^*(\underline{\Lambda}))$, with an additional constant that disappears when taking gradients. We focus on the $\log\det$ part. For this purpose, we write the block decomposition
\begin{equation*}
\left(\begin{array}{ccccc}
		L[:i-1,:i-1] + \diag^*(\underline{\Lambda}[:i-1])& \vline &  L[:i-1,i]& \vline & L[:i-1,i+1:]  \\ \hline \\
		L[i,:i-1] & \vline & \delta_i I_d+ \Lambda_i & \vline & L[i,i+1:] \\ \hline \\
		L[i+1:, :i-1] & \vline & L[i+1:,i]& \vline & L[i+1:,i+1:] + \diag^*(\underline{\Lambda}[i+1:])
	\end{array}\right). 
\end{equation*}
Moving the $i$th column of blocks to the left side and the $i$th row to the top, and exploiting that an even number of operations leaves the determinant unchanged we can write
\begin{equation*}
    \det(L + \diag^*(\underline{\Lambda})) = \det \left(
	\begin{array}{ccc}
	 \delta_i I_d + \Lambda_i & \vline & N_i^T \\
	 \hline
	N_i  & \vline & L[-i] + \diag^*(\underline{\Lambda}[-i] )
	\end{array}
	\right) \quad \text{where} \quad  N_i = I_d \otimes ({\bf 1}\{ j \sim i \})_{1 \leq j \neq i \leq p}.
\end{equation*}
In last equation $N_i \in S_d(\mathbb{R})^{p-1}$ is the tensorized neighbors vector of the $i$th vertex, $L[-i]$ is the Laplacian matrix with $i$th row and $i$th column removed, and underline $\underline{\Lambda}[-i] \in S_d(\mathbb{R})^{p-1}$ is vector $\underline{\Lambda}$ with $i$th block $\Lambda_{i}$ removed. Exploiting Schur complement formula, we have
\begin{equation*}
	\det[L+\diag^*(\underline{\Lambda})] = \det(L[-i] + \diag^*(\underline{\Lambda}[-i]))\det\left(\delta_i I_d + \Lambda_i - N_i^T(L[-i]+\diag^*\underline{\Lambda}[-i])^{-1}N_i \right).
\end{equation*}
Now, we can write
\begin{equation*}
	\log \det(L+\diag^*(\underline{\Lambda}))=\log\det(\Lambda_i + \delta_i I_d  - N_i^T(L[-i]+\diag^*\underline{\Lambda}[-i])^{-1}N_i)+ {\rm const}.
\end{equation*}
Exploiting that the gradient of $\log\det$ is given by the inverse matrix, we derive 
\begin{equation*}
	\nabla_{\Lambda_i}\log \det(L+\diag^*(\underline{\Lambda})) = (\Lambda_i + \delta_i I_d  - N_i^T(L[-i]+\diag^*\underline{\Lambda}[-i])^{-1}N_i)^{-1}. 
\end{equation*}
Reminding the linear term, the gradient of the dual function with respect to $\Lambda_i$ is given by 
\begin{equation*}
	\nabla_{\Lambda_i}\mathcal{D}(\underline{\Lambda}) = - \Sigma_i + \varepsilon (\Lambda_i + \delta_i I_d  - N_i^T(L[-i]+\diag^*\underline{\Lambda}[-i])^{-1}N_i)^{-1}.
\end{equation*}
At optimality we have $\nabla_{\Lambda_i}\mathcal{D}(\underline{\Lambda}) = 0$ which is equivalent to 
\begin{align*}
	- \Sigma_i + \varepsilon (\Lambda_i + \delta_i I_d  - N_i^T(L[-i]+\diag^*\underline{\Lambda}[-i])^{-1}N_i)^{-1} = 0 & \Leftrightarrow  \Lambda_i + \delta_i I_d  - N_i^T(L[-i]+\diag^*\underline{\Lambda}[-i])^{-1}N_i = \varepsilon \Sigma_i^{-1} \\
	& \Leftrightarrow \Lambda_i =   \varepsilon \Sigma_i^{-1}   - \delta_i I_d  + N_i^T(L[-i]+\diag^*\underline{\Lambda}[-i])^{-1}N_i.
\end{align*}
We thus have shown that at optimality, the relation 
\begin{equation}\label{eq:optima_condition_dual_block}
	 \Lambda_i =   \varepsilon \Sigma_i^{-1}   - \delta_i I_d  + N_i^T(L[-i]+\diag\underline{\Lambda}[-i])^{-1}N_i.
\end{equation}
holds true. To conclude, we need to show that $C_{\underline{\Lambda}^*}$ satisfies primal feasibility. For this purpose, we use the same block decomposition
\begin{equation*}
	[L + \diag^*(\underline{\Lambda})]^{-1} = \begin{pmatrix}
		\delta_1 I_d + \Lambda_1 & N_1\\
		N_1 & L[-1] + \diag^*\underline{\Lambda}[-1]
	\end{pmatrix}^{-1}, 
\end{equation*}
and focusing on the upper-left block, the Schur complement formula yields
\begin{equation*}
	\varepsilon[L + \diag^*(\underline{\Lambda})]^{-1} =  \begin{pmatrix}
		\varepsilon(\delta_1 I_d + \Lambda_1 - N_1^T[L[-1]+\diag^*(\underline{\Lambda}[-1])]^{-1}N_1)^{-1} & (*) \\
		(*) & (*)
	\end{pmatrix}.
\end{equation*}
And reminding relation \eqref{eq:optima_condition_dual_block} for $\Lambda_1$ enables us to rewrite the upper left block
\begin{align*}
	\varepsilon(\delta_1 I_d + \Lambda_1 - N_1^T(L[-1]+\diag^*(\underline{\Lambda}[-1]))^{-1}N_1)^{-1} & = \varepsilon(\varepsilon \Sigma_1^{-1})^{-1} \\
	& = \Sigma_1.
\end{align*}
We have shown that primal feasibility holds.
\end{proof}

\medskip

\begin{proof}[Proof of \Cref{lem:uniqueness_entropic}]
The function $-\log : [0,+\infty) \rightarrow \mathbb{R} \cup \{+ \infty\}$ is lower-semi-continuous. As a consequence, the objective function $\mathcal{F}(C) := \langle L, C\rangle_{\rm F} - \varepsilon \log \det(C)$ is lower-semi-continuous on $S_{pd}^{+}(\RR)$. As $\Pi(\underline{\Sigma})$ is compact, problem \eqref{eq:entropic-ot-gaussian} has at least one solution. If $C\in \Pi(\underline{\Sigma})$ is not positive-definite, $\mathcal{F}(C) = + \infty$. Thus, solutions to problem $\eqref{eq:entropic-ot-gaussian}$ belong to the set of positive-definite matrices $S_{pd}^{++}(\mathbb{R})$. As the $\log \det$ function is strictly concave on $S_{pd}^{++}(\mathbb{R})$ (see for example \cite[Cor.1.4.2]{bakonyi2011matrix}), the objective function of problem \eqref{eq:entropic-ot-gaussian} is strictly convex. This ensures that problem \eqref{eq:entropic-ot-gaussian} has exactly one solution. 
\end{proof}

\subsection{Omitted proofs from Section \ref{sec:identifiability}}

\begin{proof}[Proof of \Cref{prop:identifiability}]
	Let $\weights \in \cW_d$. Considering the off-diagonal blocks of KKT for \eqref{eq:entropic-ot-weighted}, $\diag^*(\underline{\Lambda})$ vanishes and $\varepsilon(C_{\weights}^{-1})_{ij}=-w_{ij}$ for $i\neq j$, i.e. $\iota(C_{\weights})=\weights$. 
	Conversely, let $C \in \Pi^{++}(\margs)$ and $w_{ij}:=-\varepsilon(C^{-1})_{ij}$ for $i\neq j$. Define $\Lambda_i:=\varepsilon(C^{-1})_{ii}-(L_{\weights})_{ii}$, then $\chi(\iota(C)) = C_{\weights} = C$. Now, note that $\iota(C)_{ij}=0$ if and only if $(C^{-1})_{ij}=0$ so $\iota$ maps bijectively $\cM(G,\margs)$ onto $\{\weights\in\cW_d:w_{ij}=0\
	\text{for}\ \{i,j\}\notin E\}$. Observing that $C_{\weights,\varepsilon}=C_{\weights/\varepsilon,1}$ gives the last remark.
\end{proof}

\begin{proof}[Proof of \Cref{prop:bias-upper}]
	Observe that $(C^{-1})_{ij}=0$ for all $i,j$ belonging to different components. Therefore, grouping by connected components, $C$ is a block diagonal matrix, and so is $C^\star$.
	Denoting $\kappa_{ij}:=\min\bigl\{\sqrt{\lambda_{\max}(\Sigma_i)\tr(\Sigma_j)}, \sqrt{\tr(\Sigma_i)\lambda_{\max}(\Sigma_j)}\bigr\}$, we get
	\begin{align*}
		\|C-C^{\star}\|_{\Fro}^{2} &= \sum_{c=1}^{m}\sum_{\substack{i,j\in V_c\\ i\neq j}} \|C_{ij}-C^{\star}_{ij}\|_{\Fro}^2 \\
		&\leq \sum_{c=1}^{m}\sum_{\substack{i,j\in V_c\\ i\neq j}}
		\bigl(\|C^{\star}_{ij}\|_{\Fro}+\kappa_{ij}\bigr)^{2} \\
		&\leq 4\sigma^4_{\max} {d \cdot \sum_{c=1}^{m}p_c(p_c-1)},
	\end{align*}
	where the last line holds since
	$\|C^{\star}_{ij}\|_{\Fro}\leq\kappa_{ij}\leq\sqrt{d}\sigma^2_{\max}$. The lower bound is given by $\diag^*(\margs)\in\cM(G, \margs)$.
\end{proof}

\subsection{Omitted results and proofs from Section \ref{sec:stability}}

\begin{lemma}[Properties of the differential]\label{lem:diff-structure}
For full-rank marginals, the linear map $\dphi:S_d(\mathbb{R})^p \to S_{dp}(\RR)$ of \eqref{eq:diff-exact} satisfies:
\begin{enumerate}
\item[\emph{(i)}] $\diag\circ \dphi=\mathrm{Id}_{\mathcal{X}}$;
\item[\emph{(ii)}] its Gram operator on $S_d(\mathbb{R})$ is
   $\dphi^{\top}\dphi=\Phi_{C^\star}^{-1}\,\Phi_{(C^\star)^2}\,\Phi_{C^\star}^{-1}$, hence
   $\|\dphi\|_{\mathrm{op}}^2=\lambda_{\max}\!\bigl(\Phi_{C^\star}^{-1}\Phi_{(C^\star)^2}\Phi_{C^\star}^{-1}\bigr)$;
\item[\emph{(iii)}] $\dphi^{\top}\dphi\succeq \Id_{\mathcal{X}}$, i.e.\ all singular values of $\dphi$ are $\ge1$.
\end{enumerate}
\end{lemma}

\begin{proof}[Proof of \Cref{lem:diff-structure}]
\emph{(i)} $\diag(\dphi(H))=\diag\bigl(C^\star\diag^{*}(\Phi_{C^\star}^{-1}H)C^\star\bigr)=\Phi_{C^\star}\bigl(\Phi_{C^\star}^{-1}H\bigr)=H$.

\emph{(ii)} Put $u:=\Phi_{C^\star}^{-1}H$: as $H$ ranges over $X$, so does $u$ ($\Phi_{C^\star}$ is invertible since $C^\star\succ0$). Then,
\[
   \|\dphi(H)\|_F^2=\|C^\star\diag^{*}(u)C^\star\|_F^2=\langle u,\Phi_{(C^\star)^2}u\rangle
   =\langle H,\Phi_{C^\star}^{-1}\Phi_{(C^\star)^2}\Phi_{C^\star}^{-1}H\rangle,
\]
the last step by self-adjointness of $\Phi_{C^\star}^{-1}$. Hence $\dphi^{\top}\dphi=\Phi_{C^\star}^{-1}\Phi_{(C^\star)^2}\Phi_{C^\star}^{-1}$, and $\|\dphi\|_{\mathrm{op}}^2=\lambda_{\max}(\dphi^\top \dphi)$ is its largest eigenvalue.

\emph{(iii)} By (i), for every $H$, $\|\dphi(H)\|_F^2\ge\|\diag(\dphi(H))\|_F^2=\|H\|_F^2$, since the diagonal blocks
are a subset of the entries of the matrix $\dphi(H)$ and Frobenius norm is monotone under restriction to a
sub-collection of entries. Thus $\dphi^\top \dphi\succeq I_X$.
\end{proof}

\medskip

\begin{proof}[Proof of \Cref{prop:bound-diff}]
	\emph{First bound.} Let $\underline{H}\in\cX$, set
	$\underline{W}:=\Phi_C^{-1}(\underline{H})$ and $M:=\diag^*(\underline{W})$,
	so that $\dphi(\underline{H})=CMC$ by \eqref{eq:diff-exact}. Writing
	$CMC=C^{1/2}\bigl(C^{1/2}MC^{1/2}\bigr)C^{1/2}$ and using
	$\|AXB\|_{\Fro}\leq\opnorm{A}\|X\|_{\Fro}\opnorm{B}$ together with
	$\opnorm{C^{1/2}}=\lambda_{\max}(C)^{1/2}$,
	\begin{equation*}
		\bigl\|\dphi(\underline{H})\bigr\|_{\Fro}
		\ \leq\ \lambda_{\max}(C)\,\bigl\|C^{1/2}MC^{1/2}\bigr\|_{\Fro}.
	\end{equation*}
	Moreover
	$\|C^{1/2}MC^{1/2}\|_{\Fro}^{2}=\tr(MCMC)
	=\langle\underline{W},\Phi_C(\underline{W})\rangle_{\cX}
	=\langle\underline{H},\Phi_C^{-1}(\underline{H})\rangle_{\cX}
	\leq\lambda_{\min}(\Phi_C)^{-1}\|\underline{H}\|_{\cX}^{2}$, since $\Phi_C$ is
	self-adjoint and positive definite. Hence
	\begin{equation}\label{eq:master-bound-intrinsic}
		\opnorm{\dphi}\ \leq\ \frac{\lambda_{\max}(C)}{\lambda_{\min}(\Phi_C)^{1/2}}.
	\end{equation}
	Moreover, observe that 
	\begin{align*}
		\bigl\langle\underline{H},\Phi_C(\underline{H})\bigr\rangle_{\cX} =\|C^{1/2}\diag^*(\underline{H})C^{1/2}\|_{\Fro}^{2}
		&\geq \lambda_{\min}(C)\,\bigl\| \diag^*(\underline{H})C^{1/2}\bigr\|_{\Fro}^{2} \\
		&=\lambda_{\min}(C)\,\tr\bigl(C\diag^*(\underline{H})^{2}\bigr) \\
		&\geq \lambda_{\min}(C) \sigma^2_{\min} \| \underline{H}\|_{\cX}^{2}.
	\end{align*}
	Hence, $\lambda_{\min}(\Phi_C) \geq \sigma^2_{\min} \lambda_{\min}(C)$ and the first bound follows.

	\emph{Second bound.}
	First, we show $\lambda_{\max}(C)\leq\lambda_{+}$. 
	On the one hand, we have the bound $\lambda_{\max}(C)\leq p\sigma^2_{\max}$. 
	On the other hand, the Schur-complement identity
	$(C^{-1})_{ii}=(\Sigma_i-C_{i,-i}
	(C_{-i,-i})^{-1}C_{-i,i})^{-1}\succeq\Sigma_i^{-1}$ gives
	$\Lambda_i^{*}=\varepsilon((C)^{-1})_{ii}-\delta_iI\succeq
	\varepsilon\Sigma_i^{-1}-\delta_iI$, so that
	$L+\diag^{*}(\underline{\Lambda}^{*})\succeq-A(G)\otimes I_d
	+\varepsilon\,\diag^{*}(\underline{\Sigma}^{-1})\succeq
	(\varepsilon/\sigma^2_{\max}-\delta_{\max})I$ and, when $\varepsilon>\delta_{\max}\sigma^2_{\max}$,
	$\lambda_{\max}(C)\le\varepsilon\sigma^2_{\max}/(\varepsilon-\delta_{\max}\sigma^2_{\max})$.
	
	It remains to show the bound on $\lambda_{\min}(C)^{-1}$.
	We have that 
	\begin{align*}
		I_d &= [C C^{-1}]_{ii} \\
		&= \sum_{1 \leq j \leq p} C_{ij} (C^{-1})_{ji} \\
		&= \Sigma_i (C^{-1})_{ii} + \sum_{1 \leq j \leq p} C_{ij} \cdot \left(- \,\frac{{\bf 1}\{\{i,j\} \in E\}}{\varepsilon} I_d \right) \\
		&= \Sigma_i (C^{-1})_{ii} - \frac{1}{\varepsilon} \sum_{j: \{i,j\} \in E} C_{ij},
	\end{align*}
	and therefore,
	\begin{align*}
		\opnorm{(C^{-1})_{ii}} &= \left\| \Sigma_i^{-1}\left( I_d + \frac{1}{\varepsilon} \sum_{j : \{i,j\} \in E} C_{ij}\right) \right\|_{\mathrm{op}}  \\
		&\leq \frac{1}{\sigma^2_{\min}} \left( 1 + \frac{1}{\varepsilon}\sum_{j : \{i,j\} \in E} \opnorm{C_{ij}}\right) \\
		&\leq \frac{1}{\sigma^2_{\min}} \left( 1 + \frac{\delta_{\max} \sigma^2_{\max}}{\varepsilon}\right) =: \alpha,
	\end{align*}
	It follows that
	\begin{align*}
		\lambda_{\min}(C)^{-1}=\lambda_{\max}(C^{-1})
		\ \leq\ \max_i\sum_j\opnorm{(C^{-1})_{ij}}
		\ \leq\ \alpha+\frac{\delta_{\max}}{\varepsilon},
	\end{align*}
	where the right hand side of the last inequality comes from the fact that  $(C^{-1})_{ij}=-\varepsilon^{-1}A(G)_{ij}I_d$ for $i\neq j$.

	Thus, we get
	\begin{align*}
		\opnorm{\varphi} &\leq \frac{\lambda_{\max}(C)}{\bigl(\sigma^2_{\min}\lambda_{\min}(C)\bigr)^{1/2}} \\
		&\leq {\lambda_{+}} \cdot \sqrt{ \frac{1}{\sigma^2_{\min}} \left(  \alpha + \frac{\delta_{\max}}{\varepsilon}\right)} \\
		&\leq \frac{\lambda_{+}}{\sigma^2_{\min}} \cdot \sqrt{1 + \frac{\delta_{\max} (\sigma^2_{\max} + \sigma^2_{\min})}{\varepsilon}}.
	\end{align*}
	
\end{proof}

\medskip

\begin{proof}[Proof of \Cref{thm:diff-graph}]
	\emph{Differential.} On
	$\cW_d\times\cZ$, with $\cZ:=S_{dp}^{++}(\RR)\times S_d(\RR)^{p}$, set
	\begin{equation*}
		F\bigl(\weights,(C,\underline{\Lambda})\bigr)
		:=\bigl(L_{\weights}-\varepsilon C^{-1}+\diag^*(\underline{\Lambda}),\
		\diag(C)-\margs\bigr).
	\end{equation*}
	Its partial differential in $(C,\underline{\Lambda})$ is the operator $\diffd f|_{\cZ}$ of \eqref{eq:differential_equation_implicit}, which was shown in the proof of \Cref{thm:regularity-implicit} to be bijective, for any $C\succ0$. 
	We have  $\partial_{\weights}F(\underline{h})=(L_{\underline h},0)$, so the implicit
	function theorem applies at every point of the zero set and $\chi$ is
	$\cC^{1}$, with $(K,\underline{V}):=\diffd(\chi,\underline{\Lambda})_{\weights}(\underline{h})$
	solving
	\begin{equation}\label{eq:linear-system-graph}
		\varepsilon C^{-1}KC^{-1}+\diag^*(\underline{V})=-L_{\underline h},
		\qquad \diag(K)=0 .
	\end{equation}
	Solving as in the proof of \Cref{thm:regularity-implicit} gives $K=-\varepsilon^{-1}C\bigl(L_{\underline h}+\diag^*(\underline{V})\bigr)C$ and with $\underline{V}=-\Phi_C^{-1}\bigl(\diag(CL_{\underline h}C)\bigr)$ yields \eqref{eq:diff-exact-graph}. 
	If $K=0$ then $L_{\underline h}=-\diag^*(\underline{V})$, whose off-diagonal blocks $-h_{ij}$ must vanish, so $\underline{h}=0$ and $\diffd\chi_{\weights}$ is one-to-one.

	\emph{Bound.} 
	Taking the inner product of \eqref{eq:linear-system-graph} with $K$ and using $\langle\diag^*(\underline{V}),K\rangle_{\Fro} =\langle\underline{V},\diag(K)\rangle_{\cX}=0$, gives, 
	\begin{equation}\label{eq:response-edges}
		\varepsilon\bigl\|C^{-1/2}KC^{-1/2}\bigr\|_{\Fro}^{2}
		=\varepsilon\tr\bigl(C^{-1}KC^{-1}K\bigr)
		=-\langle L_{\underline h},K\rangle_{\Fro}
		=\sum_{i\neq j}\langle h_{ij},K_{ij}\rangle_{\Fro}
		=2\sum_{\{i,j\}\in\cE}\langle h_{ij},K_{ij}\rangle_{\Fro}.
	\end{equation}
	We apply Cauchy--Schwarz inequality to bound the right-hand side by
	$\sqrt{2}\|\underline{h}\|_{\cW_d}\|K\|_{\Fro}$, while
	$K=C^{1/2}(C^{-1/2}KC^{-1/2})C^{1/2}$ and
	$\|AXB\|_{\Fro}\leq\opnorm{A}\|X\|_{\Fro}\opnorm{B}$ bound the left-hand side
	below by $\varepsilon\|K\|_{\Fro}^{2}/\lambda_{\max}(C)^{2}$. Hence
	$\|K\|_{\Fro}\leq\sqrt{2}\lambda_{\max}(C)^{2}\varepsilon^{-1}\|\underline{h}\|_{\cW_d}$.
	Finally $\lambda_{\max}(C_{\weights})\leq\lambda_{+}$ as in \Cref{prop:bound-diff}.
\end{proof}


\subsection{Omitted results and proofs from Section \ref{sec:statistical-scenario}}\label{subsec:proofs_inference}

\begin{lemma}\label{lem:gauss-scale}
	Let $Z\sim N(0,K)$ take values in a finite-dimensional Euclidean space, $K\neq0$. Then
	$\sqrt{\frac{2}{\pi}\tr K}\le\mathbb E\|Z\|\le\sqrt{\tr K}$.
\end{lemma}

\begin{proof}[Proof of \Cref{lem:gauss-scale}]
	The upper bound is a consequence of Jensen's inequality.
	For the lower bound, write $K=U\Lambda U^\top$ with
	$\Lambda=\mathrm{diag}(\lambda_1,\dots,\lambda_d)$, $\lambda_i\ge0$,
	and let $Y=(y_1,\dots,y_d)\sim N(0,I_d)$. Then, in distribution we have
	$Z= U\Lambda^{1/2}Y$ and
	$\|Z\| = (\sum_{i=1}^d \lambda_i y_i^2)^{1/2}$.
	Set $s=\operatorname{tr}K=\sum_i\lambda_i$ and $a_i=\lambda_i/s$. Define
	\begin{align*}
		F(a_1,\dots,a_d)
		:=\mathbb E\Bigl(\sum_{i=1}^d a_i y_i^2\Bigr)^{1/2},
		\qquad \sum_{i=1}^d a_i=1,\ a_i\ge0.
	\end{align*}
	As a function of $(a_1,\ldots,a_d)$, $F$ is concave on the simplex. Therefore $F$ attains its minimum at an extreme point, say $e_1$:
	$F(a)\ge F(e_1)=\mathbb E|y_1|=\sqrt{\frac{2}{\pi}}$.
	Consequently,
	\begin{align*}
		\mathbb E\|Z\|
		=\sqrt{s}\,F(a)
		\geq \sqrt{\frac{2}{\pi}}\,\sqrt{\operatorname{tr}K}.
	\end{align*}
\end{proof}

\begin{proof}[Proof of \Cref{thm:nonasymptotic}]
    The first inequality is straightforward by \Cref{thm:regularity-implicit} and by convexity $\cK(\sigma^2_{\min}, \sigma^2_{\max})$. By assumption \ref{assum:sampling} and \cite[Thm.~4.7.1]{vershynin2018high}, there is a constant $c_0>0$ such that, for each $i$ and each $u>0$, with probability at least $1-2e^{-u}$,
    \begin{equation*}
        \opnorm{\widehat{\Sigma}_i-\Sigma_i^{\star}}
        \ \leq\ c_0\,\opnorm{\Sigma_i^{\star}}
        \max\Bigl\{\sqrt{\tfrac{d+u}{n_i}},\ \tfrac{d+u}{n_i}\Bigr\}.
    \end{equation*}
    Taking $u=\log(2p/\alpha)$ and a union bound over $i\leq p$, with probability at least $1-\alpha$ this holds for all $i$ simultaneously; on that event, using $n_i\geq\gamma_{\min}n$ and
    $\opnorm{\Sigma_i^{\star}}\leq\sigma^{2}_{\max}$, and for $n$ large enough,
    \begin{equation}\label{eq:dev-half}
        \max_i\opnorm{\widehat{\Sigma}_i-\Sigma_i^{\star}}
        \ \leq\ c_0\,\sigma^{2}_{\max}\sqrt{\tfrac{d+u}{\gamma_{\min}n}}
        \ \leq\ \tfrac12\,\sigma^{2}_{\min}.
    \end{equation}
    It follows that both $\widehat{\margs}_n$ and $\margs^{\star}$ lie in $\cK(\tfrac12\sigma^{2}_{\min},2\sigma^{2}_{\max})$. Using $\|A\|_{\Fro}\leq\sqrt{d}\opnorm{A}$ blockwise, we obtain 
    \begin{equation*}
        \bigl\|\widehat{C}_n-C^{\star}\bigr\|_{\Fro}
        \ \leq\ K(\varepsilon, p, \delta_{\max}) \bigl\|\widehat{\margs}_n-\margs^{\star}\bigr\|_{\cX}
        \ \leq\ K(\varepsilon, p, \delta_{\max}) \sqrt{pd}\,
        \max_i\opnorm{\widehat{\Sigma}_i-\Sigma_i^{\star}},
    \end{equation*}
    which concludes the proof.
\end{proof}

\begin{proof}[Proof of \Cref{thm:clt}]
	
\emph{Central limit Theorem.} This corresponds exactly to the multivariate delta method.
By \Cref{thm:regularity-implicit}, $\varphi$ is continuously differentiable on an open
neighborhood $\Omega$ of $\underline{\Sigma}^\star$; in particular its first component
$\varphi$ is differentiable at $\underline{\Sigma}^\star$, with differential
$\diffd\varphi_{\underline{\Sigma}^\star}\in\mathcal L\bigl(S_d(\RR)^p,\,S_{dp}(\RR)\bigr)$,
both spaces being endowed with the Frobenius inner product.
The assumption $a_n(\widehat{\underline{\Sigma}}_n-\underline{\Sigma}^\star)\xrightarrow{d} N(0,\Gamma)$ with
$a_n\to\infty$ implies $\widehat{\underline{\Sigma}}_n\xrightarrow{\mathbb P}\underline{\Sigma}^\star$ (a sequence
converging in distribution is tight, and $a_n^{-1}\to0$). Hence
$\mathbb P(\widehat{\underline{\Sigma}}_n\in\Omega)\to1$, so $\varphi(\widehat{\underline{\Sigma}}_n)$ is well
defined with probability tending to one.
Applying \cite[Theorem.~3.1]{van2000asymptotic} gives the result.\\

\emph{Rate in probability.} By the continuous mapping theorem applied to the Frobenius norm, $\sqrt n\|\widehat C_n-C^\star\|_{\rm F} \xrightarrow{d}\|G\|_{\rm F}$.
As a sequence converging in distribution is tight, we have $\sqrt n\|\widehat C_n-C^\star\|_{\rm F}=O_{\PP}(1)$.
By \Cref{lem:gauss-scale}, it follows that $\|\widehat C_n-C^\star\|_{\rm F}=O_{\PP}(\sqrt{T/n})$.
Moreover, for any $\delta > 0$, there is $c>0$ such that $\PP(\|G\|_{\rm F}>c)\geq 1-\delta$ (because $G$ is non-degenerate). 
By the Portmanteau theorem on the open set $\{x>c\}$, $\liminf_n\PP(\sqrt n\|\widehat C_n-C^\star\|_{\rm F}>c)\geq \PP(\|G\|_{\rm F}>c)\geq 1-\delta$; hence
$\|\widehat C_n-C^\star\|_{\rm F}$ is not $o_{\PP}(\sqrt{T/n})$.

Since $\Gamma\succeq0$, $T=\tr(\Gamma^{1/2}\dphi^\top \dphi\,\Gamma^{1/2})$ lies between
$\lambda_{\min}(\dphi^\top \dphi)\tr\Gamma$ and $\lambda_{\max}(\dphi^\top \dphi)\tr\Gamma$. Equality holds when $\Gamma^{1/2}$
projects onto the corresponding eigenvector of $\dphi^\top \dphi$. 
\end{proof}

\begin{proof}[Proof of \Cref{prop:exact-epsilon}]
    The $\{i,j\}$ block of $(C^{\star})^{-1}$ is the
    inverse of the conditional covariance of $(X_i,X_j)$ given the remaining variables.
    It follows that
    \begin{equation*}
        (C^{\star})^{-1}_{ij}
        = -\bigl(\Sigma^{|}_{ii}-\Sigma^{|}_{ij}(\Sigma^{|}_{jj})^{-1}
        (\Sigma^{|}_{ij})^{\top}\bigr)^{-1}\Sigma^{|}_{ij}(\Sigma^{|}_{jj})^{-1}.
    \end{equation*}
    Moreover, by the KKT conditions $(C^{\star})^{-1}_{ij} = -\varepsilon^{-1}I_d$. Hence
    \begin{equation*}
        \varepsilon\,\Sigma^{|}_{ij}(\Sigma^{|}_{jj})^{-1}
        = \Sigma^{|}_{ii}-\Sigma^{|}_{ij}(\Sigma^{|}_{jj})^{-1}(\Sigma^{|}_{ij})^{\top},
    \end{equation*}
    and taking the trace gives the result.
\end{proof}

\end{document}